\documentclass[useAMS,usenatbib]{mn2e}
\usepackage{graphicx}
\usepackage{enumerate}
\usepackage{longtable,lscape}
\usepackage{rotating}
\usepackage{multirow,multicol}

\usepackage[usenames,dvipsnames]{xcolor}
\usepackage[paperheight=297mm,paperwidth=210mm,margin=0.7in]{geometry}
\usepackage[T1]{fontenc} 
\usepackage{aecompl}

\title[The properties of early-type galaxies in the Ursa Major cluster]{The properties of early-type galaxies in the Ursa Major cluster}

\author[Mina Pak et al.]{Mina Pak$^{1}$\thanks{E-mails: minapak.astro@gmail.com; screy@cnu.ac.kr  (corresponding authors)}, Soo-Chang Rey$^{1}$$^{\star}$, Thorsten Lisker$^{2}$, Youngdae Lee$^{1}$, Suk Kim$^{1}$, \newauthor Eon-Chang Sung$^{3}$, Helmut Jerjen$^{4}$, and Jiwon Chung$^{1}$\\
$^{1}$Department of Astronomy and Space Science, Chungnam National University, 99 Daehak-ro, Daejeon 305-764, Korea\\
$^{2}$Astronomisches Rechen-Institut, Zentrum f\"ur Astronomie der Universit\"at Heidelberg, M\"onchhofstra\ss e 12-14, 69120 Heidelberg, Germany\\
$^{3}$Korea Astronomy and Space Science Institute, Daejeon 305-348, Korea\\
$^{4}$Research School of Astronomy and Astrophysics, The Australian National University, Cotter Road, Weston, ACT, 2611, Australia}

\pagerange{\pageref{firstpage}--\pageref{lastpage}} \pubyear{2014}

\begin{document}
\label{firstpage}
\maketitle

\begin{abstract}
Using SDSS-DR7 and NASA/IPAC Extragalactic Database spectroscopic data, we identify 166 galaxies as members of the Ursa Major cluster with M$_{r}$ $<$ $-13.5$ mag. We morphological classify all galaxies by means of carefully inspecting $g-$, $r-$, $i-$band colour and monochromatic images. We show that the Ursa Major cluster is dominated by late-type galaxies, but also contains a significant number of early-type galaxies, particularly in the dwarf regime. We present further evidence for the existence of several subgroups in the cluster, consistent with previous findings. The early-type fraction is found to correlate with the mass of the subgroup. We also investigate environmental effects by comparing the properties of the Ursa Major early-type dwarf galaxies to those of the Virgo cluster. In contrast to the Virgo, the red sequence of the Ursa Major cluster is only sparsely populated in the optical and ultraviolet colour-magnitude relations. It also shows a statistically significant gap between $-18$ $<$ M$_{r}$ $<$ $-17$ mag, i.e.\ the Ursa Major cluster lacks early-type dwarf galaxies at the bright end of their luminosity function. We discover that the majority of early-type dwarf galaxies in the Ursa Major cluster have blue cores with hints of recent or ongoing star formation. We suggest that gravitational tidal interactions can trigger central blue star forming regions in early-type dwarfs. After that, star formation would only fade completely when the galaxies experience ram pressure stripping or harassment,
both of which are nearly absent in the Ursa Major cluster.

\end{abstract}


\section{INTRODUCTION}
Galaxy properties, such as morphological appearance and star formation rate (SFR), strongly depend on the physical conditions of the environment.
Clusters of galaxies are dominated by red early-type galaxies with little star-forming activity, while the dominant population in the low-density field are blue late-type galaxies with significant star formation (\citealt{Dre80}; \citealt{Bin87}; \citealt{Lew02}). 
 
The trend that the fraction of early-type galaxies increases with the local density is found even in poor groups (\citealt{Pos84}; \citealt{Fer91}; \citealt{Dre97}). Cosmological simulations predict that galaxies and groups are assembled into massive clusters moving along filamentary structures (e.g.\ Millennium Simulation; \citealt{Spr05}). These surrounding regions are likely to play a critical role in the evolution of galaxies before becoming cluster galaxies. It is reported that ultraviolet-optical colours change sharply from blue to red when going from the field to dense environments (\citealt{Kod01}; \citealt{Tan05}), implying that at least for some of the red cluster galaxies, star-forming activity was quenched through environmental effects before entering the cluster region. Moreover, galaxy groups, due to their low velocity dispersion, represent natural sites for the previous evolution of present-day cluster galaxies through tidal interactions \citep{Mam90}, before the galaxies settled down in clusters. It is thus likely that at least part of the morphological transformation of cluster galaxies took place in earlier epochs in different conditions than the ones observed in present day clusters \citep{Lis13}. This is called pre-processing --- its relevance has still been debated in recent simulation results (\citealt{Zab98}; \citealt{McG09}; \citealt{DeL12}).

There is a range of known physical mechanisms that could be responsible for altering galaxy morphology and/or SFR in clusters. Ram pressure stripping \citep{Gun72}, one of the most powerful effects, seems to have been observed directly in the Virgo cluster \citep{Chu07}, and simulations suggest that this process operates on short time-scales ($\sim$1 Gyr, \citealt{Roe09}). Other processes include strangulation (\citealt{Lar80}; \citealt{Kau93}), which cuts off the cold gas supply for ongoing star formation in cluster galaxies. Galaxy mergers, tidal interactions, and harassment within the cluster potential \citep{Moo96} are other effective processes in transforming observed properties of the galaxy populations.
  
When investigating the effect of the environment on the evolution of galaxies, early-type galaxies and transitional dwarf galaxies in poor groups are the most attractive targets. In general, dwarf galaxies with a shallow potential well are more sensitive to environmental effects than giant galaxies. Recent studies reveal blue-cored early-type dwarf galaxies in various environments: clusters (\citealt{Lis06}; \citealt{DeR03}, \citealt{DeR13}), groups (\citealt{Tul08}; \citealt{Cel05}), and the field \citep{Gu06}. Theories recently evoked include that blue-cored early-type dwarf galaxies are the results of morphological evolution of late-type star forming galaxies by cluster environmental effects such as ram pressure stripping, harassment, tidal interaction or tidal stirring, and mergers (\citealt{May01a}; \citealt{Lis06}; \citealt{Lis09}). It is also possible that blue-cored early-type dwarf galaxies in fields are formed via merging of gas rich small irregular galaxies \citep{Bek08}. While all these hypotheses are still debatable, comparing the presence and properties of these transitional dwarf galaxies in different environments is thus a promising approach to shed light on the specific transformation mechanisms.
  
In this context, the Ursa Major cluster is an ideal target to investigate the evolution of dwarf galaxies. The cluster is located at a distance of 17.4 Mpc \citep{Tul12}, slightly more distant than the Virgo cluster (15.8 Mpc, \citealt{Jer04}; 15.9 Mpc, \citealt{Tul12}) but closer than the Fornax cluster (20.3 Mpc, \citealt{Jer03}). \citet{Tul96} mentioned that the Ursa Major cluster is situated in the junction of large-scale filamentary structures, which makes it intrinsically difficult to measure its dynamical properties and to define cluster membership for galaxies in this region. To overcome this problem, \citet{Tul96} defined the cluster members to be galaxies within a projected circle with 7$\degr$.5  radius from the nominal cluster centre (RA(B1950) = 11$^h$ 56$^m$.9 and Dec.(B1950) = $+$49$\degr$ 22$\arcmin$) by means of a hierarchical grouping algorithm. The Ursa Major cluster covers the velocity range between 700 and 1210 km s$^{-1}$ and has a small velocity dispersion of 148 km s$^{-1}$ \citep{Tul96}. \citet{Tul96} identified 79 cluster members which consisted mainly of gas-rich late-types \citep{Ver01a} and two dozen dwarf galaxies \citep{Tre01}. Morphological and photometric properties in the optical, near-infrared, and radio of individual galaxies are described in detail in \citet{Tul96} and \citet{Ver01a}.
 
In the following, we present a photometric study of the optical and ultraviolet (UV) properties of galaxies in the Ursa Major cluster. We compile previously uncatalogued cluster galaxies using the spectroscopic data of the Sloan Digital Sky Survey (SDSS) and the UV data of the Galaxy Evolution Explorer ($GALEX$).
By comparing their properties to those in the more massive and more dense Virgo cluster, we examine  different environmental effects on galaxy properties. This paper is arranged as follows: section 2 describes the criteria for the selection of the cluster galaxy samples and how the photometry is performed. In section 3, the galaxy catalogue is listed and the properties of the Ursa Major cluster are presented in comparison to previous studies. In section 4, we show our results on the galaxy properties and compare them to the Virgo cluster. A discussion on the environmental dependence is given in section 5, along with our conclusions.

\section{DATA AND ANALYSIS}
\subsection{SDSS data and selection of galaxies} 
\begin{figure}
\centering
\includegraphics[scale=0.4]{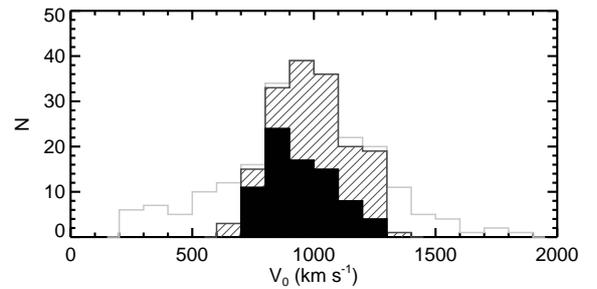}
\caption{Radial velocity distribution of galaxies in the Ursa Major cluster included in a projected circle with 7$\degr$.5 radius (cf.\ \citealt{Tul96}). The hatched histogram denotes our 166 member galaxies. The grey histogram is for the galaxies located in the same circle, but not  selected as member galaxies. The filled histogram is for the 79 member galaxies of \citet{Tul96}, i.e.\ the 12-1 group from \citet{Tul87}.}\label{F1}
\end{figure}

\begin{figure}
\centering
\includegraphics[scale=0.4]{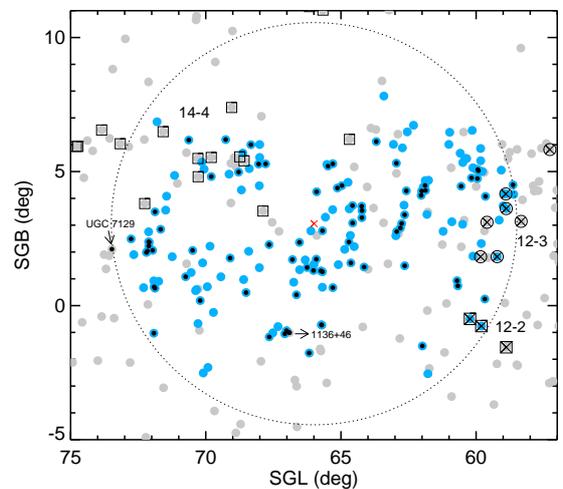}
\caption{Projected spatial distribution of galaxies in the Ursa Major cluster plotted in Supergalactic coordinates.  The dotted large circle with 7$\degr$.5 radius centred at R.A.(B1950) = 11$^h$ 56$^m$.9 and Dec.(B1950) = $+$ 49$\degr$ 22$\arcmin$ (SGL = 66$\degr$.03 and SGB = 3$\degr$.04, red cross) defines the region of the Ursa Major cluster \citep{Tul96}. All galaxies extracted from the SDSS and NED with V$_{0}$ $<$ 2000 km s$^{-1}$ are presented by grey filled circles. Blue filled circles denote our 166 cluster member galaxies whereas black filled circles are the original 79 member galaxies from the \citet{Tul96} catalogue. Galaxies in neighbouring groups defined by \citet{Tul87} are also indicated:  open squares for the 14-4 group, open squares with cross for the 12-2 group, and open circles with cross for the 12-3 group. The galaxies 1136+46 and UGC 7129 from \citet{Tul96} are excluded from our catalogue; 1136+46 is a background galaxy with a large SDSS redshift and UGC 7129 is just located outside of the 7$\degr$.5 radius circle.}\label{F2}
\end{figure}

Our sample in the Ursa Major cluster is predominantly based on the SDSS Data Release 7 (DR7; \citealt{Aba09}). The SDSS imaging data provide reduced and calibrated CCD images taken in the $u$, $g$, $r$, $i$, and $z$ bands, which allow detailed morphological classification and accurate galaxy photometry. We took advantage of the SDSS spectroscopic data for determining cluster membership of galaxies.

Following the definition of the region of the Ursa Major cluster by \citet{Tul96}, we extract SDSS spectroscopic data for all galaxies within a 7$\degr$.5 distance from the cluster centre at RA (J2000) = 11$^h$ 59$^m$ 28$^s$.4 and Dec.(J2000) = $+$49$\degr$ 05$\arcmin$ 18$\arcsec$.0. As a first step, we selected 162 galaxies with radial velocities less than 2000 km s$^{-1}$ in this region. While the spectra of most SDSS galaxies are obtained from the fibre aperture near the centre of galaxies, some large galaxies have multiple spectra observed at various fibre locations. In this case, we choose the radial velocity derived from the spectrum obtained at the nearest location to the galaxy centre.

The SDSS spectroscopic observations cover galaxies as faint as r = 17.77 mag \citep{Str02}, but luminous galaxies were rejected in the SDSS spectroscopic observations due to saturation problems (Blanton et al. 2005a, b). In order to fill in the incompleteness, we added 69 galaxies from the NASA/IPAC Extragalactic Database (NED), which are located in the same region and have the same radial velocity range. The additional galaxies from NED cover a wide range of magnitude (9.8 $<$ r $<$ 17.7), but are dominated by galaxies brighter than r = 14 mag.

From the radial velocity distribution of the galaxies, we determined Ursa Major cluster membership using the biweighted scale estimator \citep{Bee90}, which provides a statistically robust mean velocity and dispersion. Galaxies with radial velocities beyond 2$\sigma$ from the mean value were rejected, and the process was repeated iteratively until convergence. 

This identifies 166 member galaxies in the Ursa Major cluster. Their velocity distribution is given in Fig. \ref{F1} (see hatched histogram).  Following \citet{Tul96}, we calculated Local Group velocity V$_{0}$ = V$_{helio}$ + 300sin\textit{l}cos\textit{b}, where \textit{l} and \textit{b} is Galactic longitude and Galactic latitude, respectively. The resulting velocity range and velocity dispersion of our sample galaxies are  677 $<$ V$_{0}$ $<$ 1302 km s$^{-1}$ and $\sigma_{V_{0}}$ = 158 km s$^{-1}$. These values are in good agreement with the results for the \citet{Tul96} galaxies (i.e.,  700 $<$ V$_{0}$ $<$ 1210 km s$^{-1}$ and $\sigma_{V_{0}}$ = 148 km s$^{-1}$, see filled histogram in Fig. \ref{F1}). 

We calculated the number ratio of the member galaxies from SDSS spectroscopic data to the total sample (NED and SDSS) as a function of magnitude. Almost all faint galaxies are included in the SDSS spectroscopic sample ($\sim$ 96 per cent at r $>$ 14 mag), while only  63 per cent of galaxies brighter than r = 12 mag are included. As we already mentioned, this is because bright galaxies with r $<$ 14.5 were omitted in the SDSS spectroscopic survey due to saturation problems (Blanton et al. 2005a, b). Additionally, fibre conflicts affect the completeness \citep{McI08}.
In order to check for any potential members of the Ursa Major cluster that we might have missed, we used the measurements from the SDSS photometric pipeline to extract galaxies within the same magnitude, size, and colour range: $r-$band Petrosian magnitude between 9 and 18, 3.4 $<$ Petro R50 $<$ 7 arcsec, and $-0.1$ $<$ $g-r$ $<$ 1.0 (using Petrosian $g-$ and $r-$band magnitudes), which corresponds to the member galaxies of the Ursa Major cluster. This yields a sample of 270 objects classified as galaxies without any spectroscopic information by the SDSS pipeline. We visually inspected every object and excluded it if it was a star, a `piece' of a galaxy instead of a whole galaxy, or a small object that is obviously located at larger distance. Most of the 270 objects turned out to be stars or background galaxies. We found 10 galaxies that bear some similarity to low surface brightness (LSB) members of the Ursa Major cluster and could not be excluded as cluster member candidates. We thus conclude that our above sample of 166 member galaxies is complete to more than 94 per cent within 9 $<$ r $<$ 18 magnitude.

The spatial distribution of galaxies selected as our final sample of the Ursa Major cluster is shown in Fig. \ref{F2} (see blue filled circles). In comparison, the 79 members defined by Tully et al. (1996, i.e., 12-1 group from \citealt{Tul87}) are also overplotted (black filled circles). All but two galaxies (1136+46 and UGC 7129) of the \citet{Tul96} sample are included in our catalogue. The galaxy 1136+46 has a large redshift of z $\sim$ 0.05 in the SDSS spectroscopic measurement. The galaxy UGC 7129 is located at 7$\degr$.53 from the cluster centre, just outside of the 7$\degr$.5 radius circle that defines our catalogue and is therefore not included. Our final catalogue thus remains at a total number of 166 galaxies.

We note that all galaxies in the foreground 14-4 group (open squares) in \citet{Tul87} are excluded from our catalogue. However, there is possibly a little contamination of background groups in the spatial and velocity distributions, as already mentioned by \citet{Tul96}. Five galaxies in our sample overlap in velocity with those in background groups in \citet{Tul87}; two galaxies in the 12-2 group (open squares with cross) and three galaxies in the 12-3 group (open circles with cross). We note that none of our result depends on the inclusion of these galaxies. The projected spatial distribution of the new Ursa Major cluster galaxies shows essentially a diffuse configuration and lack of concentration towards any core. However, the new member galaxies in our catalogue seem to fill in the vacant regions between the \citet{Tul96} cluster members. Consequently, the overall distribution of galaxies appears to be a more elongated, filamentary configuration in the direction of supergalactic longitude.

\subsection{$GALEX$ ultraviolet data} 
\begin{figure}
\centering
\includegraphics[scale=0.7]{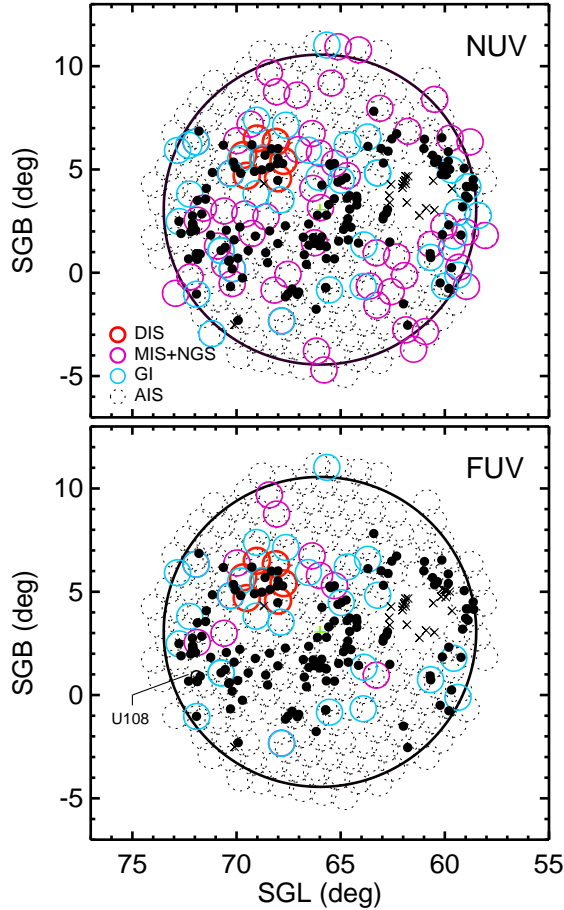}
\caption{Projected spatial distribution of $GALEX$ NUV (top panel) and FUV (bottom panel) detected galaxies in Supergalactic coordinates. Crosses denote galaxies (18 and 25 in NUV and FUV, respectively) which are not covered by $GALEX$ observations. The small open circle in the lower panel indicates U108, which is observed but not detected in FUV. Open circles with different colours imply $GALEX$ mosaic fields with different imaging modes. The solid large circle with 7$\degr$.5 radius indicates the Ursa Major cluster region based on \citet{Tul96}.} \label{F3}
\end{figure}

In order to investigate the ultraviolet properties of our sample galaxies in the Ursa Major cluster, we used far-ultraviolet (FUV; 1350 $-$ 1750 \AA) and near-ultraviolet (NUV; 1750 $-$ 2750 \AA) images from the $GALEX$ GR6. A $GALEX$ pointing has a circular field of view of 1.$\degr$25 diameter \citep{Mor07}. In order to secure the best astrometric and photometric results we avoided the image distortions at the edge of the field and considered galaxies located only within the inner 0.55 degree radius field. The $GALEX$ plate scale is 1.5 arcsec pixel $^{-1}$ which adequately samples the  4.5 $-$ 6 arcsec point spread function. 

$GALEX$ observed a total of 316 and 263 fields mostly overlapping in NUV and FUV, respectively, within a 7$\degr$.5 radius circle from the centre of the Ursa Major cluster (see Fig. \ref{F3} for mosaic fields for NUV and FUV). Unfortunately, due to some coverage gaps 18 and 25 galaxies were missed in NUV and FUV, respectively (see crosses in Fig. \ref{F3}).
 The depth of each field also varies with the survey mode: Deep Imaging Survey (DIS), Medium Imaging Survey (MIS), Nearby Galaxy Survey (NGS), All-sky Imaging Survey (AIS), and Guest Investigator (GI) programme. When several UV images are available for a galaxy, we used the deepest image for its UV source. Overall, 148 ($\sim$89 per cent) and 140 ($\sim$84 per cent) galaxies were detected in NUV and FUV, respectively. This means that all observed galaxies were detected NUV sources and only a single one of those galaxies that were covered remained undetected in FUV (U108 in Fig. \ref{F3}). All FUV detected galaxies are also detected in the NUV. In Table 1, we summarize the $GALEX$ observations and the number of detected galaxies in different survey modes.

\begin{table}
\caption{Properties of $GALEX$ data}
\begin{center}
\scriptsize
\begin{tabular}{@{}lccc}
\hline \hline
Mode & Observed field \# & Exposure time (s) & Detected galaxy \#\\
\hline\hline
   \multicolumn{4}{c}{NUV}\\
\hline
DIS	& 7	& 8969 - 13756 & 12\\
MIS	& 50	 & 126 - 2982 &	-\\
NGS	& 6	& 109 - 1242 & 9\\
AIS	& 216 & 71 - 856 & 102\\
GI	& 37 & 326 - 9482 & 25\\
\hline
   \multicolumn{4}{c}{FUV}\\
\hline
DIS 	& 7 & 8969 - 13756 &	12\\
MIS 	& 7 & 1115 - 2026 & -\\
NGS 	& 6 & 107 - 1242 & 9\\
AIS & 216 & 71 - 409 & 103\\
GI &	 27 & 1444 - 5891 & 16\\

\hline
\end{tabular}
\end{center}
\end{table}

\subsection{Optical and ultraviolet photometry} 
The SDSS photometric pipeline is optimized for small, faint objects. Galaxies with large angular extent, irregular morphology, and bright subclumps (e.g., HII regions and spiral arms) are shredded into multiple, separate objects. The standard SDSS pipeline thus fails to derive a reliable total flux of these galaxies \citep{Lis07}. In the SDSS pipeline, a 256 $\times$ 256 pixel mask was used for determining the sky background. However, large and/or bright galaxies -- in terms of apparent size and magnitude -- occupy a considerable fraction of the 256 $\times$ 256 pixel mask, resulting in an overestimation of sky value as high as 1 magnitude (see \citealt{Wes10}). In order to avoid these two issues, we performed our own SDSS photometry for all selected galaxies in the Ursa Major cluster.

The photometry of member galaxies was performed using Source Extractor (SE{\tiny{XTRACTOR}}; \citealt{Ber96}). For each galaxy, we created SDSS postage images in five passbands ($u$, $g$, $r$, $i$, and $z$). Source detection and deblending are carried out on the r-band image. The same source position and isophote of r-band image are also used for subsequent photometry in other passbands. We use the parameter BACK SIZE = 256 pixels for most galaxies. In the case of large, extended galaxies, a larger mesh size (i.e., BACK SIZE = 512 pixels) is adopted to prevent an overestimation of the sky background. For source detection, we usually chose parameters of DETECT THRESHOLD = 1 and DETECT MINAREA = 10. In some cases of faint galaxies and galaxies affected by neighbouring bright sources, different parameters (DETECT THRESHOLD = 0.5 $\sim$ 1, DETECT MINAREA = 5 $\sim$ 50) were chosen preventing a large number of spurious detections. The deblending parameters (DEBLEND MINCONT and DEBLEND NTHRESH) for each galaxy were selected iteratively by visual inspection of detected sources, enabling to separate neighbour sources which overlap with the target galaxy. We adopted MAG AUTO as the total magnitude of a galaxy, which provides the flux within an adjustable elliptical aperture that is scaled according to the isophotal light profile. Once a flux was calculated, we converted all fluxes to asinh magnitudes \citep{Lup99}. As for the $GALEX$ UV sources, we also performed photometry for all detected objects in a similar manner. Flux calibrations were applied to bring the final photometry to the AB magnitude system \citep{Oke90}. 

Magnitude errors of SDSS photometry were estimated using the equations given by \citet{Lis08} in which several different sources of uncertainty are considered (see Sec. 4 of \citealt{Lis08} for details). Magnitude errors of $GALEX$ photometry were estimated by adding the three different uncertainties in quadrature: photometric calibration uncertainty, signal-to-noise (S/N) uncertainty, and background uncertainty. The calibration uncertainty is estimated to be 0.03 and 0.05 mag in the NUV and FUV, respectively \citep{Mor07}. The S/N uncertainty is calculated using the equations given by the $GALEX$ Data Guide on the MAST homepage\footnote{http://galexgi.gsfc.nasa.gov/docs/galex/Documents/GALEXPipelineDataGuide.pdf}. SE{\tiny{XTRACTOR}} assumes that the distribution of background counts is a Gaussian distribution. However, due to the low count rates in the UV images, a Poisson distribution is a better model for the background count rate. To account for this effect, we used the equation provided by the $GALEX$ Data Guide,

\begin{displaymath}  dm = 1.086 \left( \frac{\sqrt{(f+s\Omega)t}}{ft} \right),\end{displaymath}

where \textit{f} is the flux of the source in counts s$^{-1}$, \textit{s} is the sky level in counts s$^{-1}$ pixel$^{-1}$, \textit{$\Omega$} is the area in which the flux is measured, and \textit{t} is the effective exposure time in seconds. The sky levels were measured from background images provided on the GALEX website. The typical values of uncertainty are 0.03 and 0.07 mag in the NUV and FUV, respectively. Colour errors were obtained by adding the relative flux errors in quadrature in each band.

The Galactic extinction correction for each galaxy is applied using the reddening maps by \citet{Sch98}.  We used the reddening law of \citet{Car89} to derive the following: R$_{FUV}$ = 8.16, R$_{NUV}$ = 8.90, R$_{u}$ = 5.05, R$_{g}$ = 3.89, R$_{r}$ = 2.82, R$_{i}$ = 2.18, and R$_{z}$ = 1.57.

\section{The Ursa Major cluster catalogue}
\subsection{Galaxy morphology}
\begin{figure*}
\centering
\includegraphics[scale=1.]{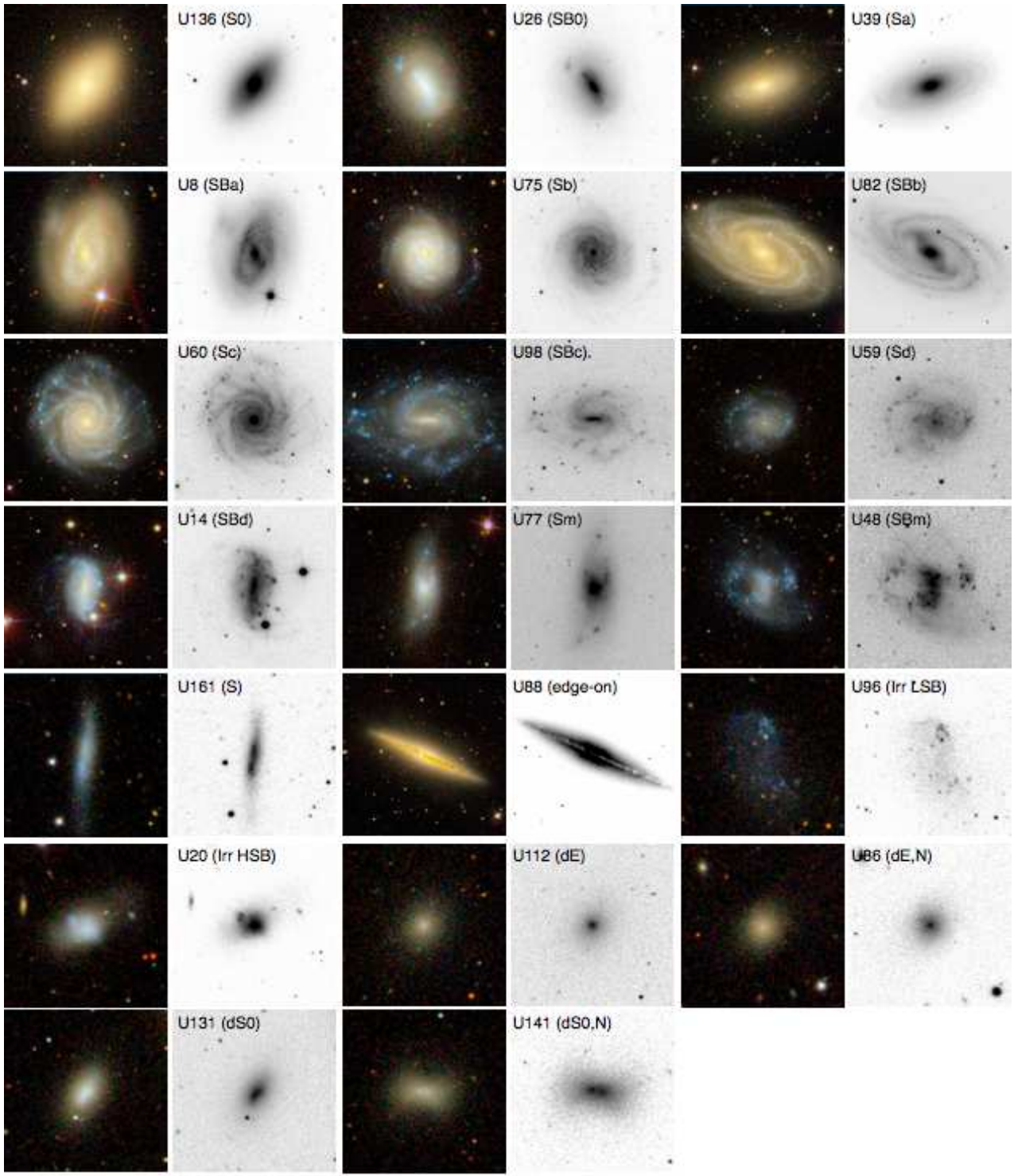}
\caption{Examples of galaxies from the different morphological classes. For each galaxy, the left- and right-hand panel shows the SDSS $g-$, $r-$, $i-$combined colour image and the $r-$band grey-scale image, respectively. ID number and morphology of each galaxy are also indicated in the upper left part of the $r-$band image.}\label{F4}
\end{figure*}

All galaxies of our catalogue were morphologically classified using the SDSS images. Five members of our team (MP, TL, SK, ES, and HJ) independently carefully inspected both monochromatic $g-$, $r-$, $i-$band images and $g-$, $r-$, $i-$combined colour images. The morphology of each galaxy was finalized if the classification of three or more classifiers agreed. If the classification for a galaxy differed for more than three classifiers, all classifiers inspected the galaxy again and agreed on a final classification.

Galaxies were divided into four major classes: early-type galaxies [elliptical (E) and lenticular (S0)], early-type dwarf galaxies [dwarf elliptical (dE) and dwarf lenticular (dS0)], spiral galaxies (Sa, Sb, Sc, and Sd), and irregular galaxies. Our morphological classification scheme is basically in accordance with the Hubble classification (see \citealt{But13} for a recent reviews). E galaxies are characterized by high surface brightness (HSB) spheroidal shape and a smooth featureless envelope with no sign of a disc. S0 galaxies have a smooth bulge and a disc component with a distinct change in surface brightness at the transition radius, discriminating them from E galaxies. In usual cases, S0 galaxies also exhibit lens-like features. The classification of spiral galaxies is based on the prominence of bulge and the appearance of the spiral arms such as degree of their tightness and resolution. Edge-on galaxies are spiral galaxies that are seen edge-on, preventing us from a more detailed morphological classification. The `S' type includes spiral galaxies for which the spiral subclass could not be determined due to relatively small angular size and lack of unique features within the spiral galaxy classification scheme. Sm galaxies are defined as Magellanic type spirals. As for the S0 and spiral galaxies, we also considered barred galaxies showing a linear bar structure in the centre.

In addition, we classify specific dwarf galaxies as described by \citet{San84}. dE galaxies are LSB ones characterized by elliptical isophotes and smooth surface brightness profiles. dS0 galaxies show an overall smooth structure but exhibit additional features deviated from the normal dE galaxies;  a distinct bulge-disc transition, asymmetric features (e.g., bar, lense, and boxyness), or an irregularity in the central region (see \citealt{Bin91} for details). In most cases of dS0 galaxies with central irregularity, their irregular structures turned out to be smaller than half-light radius returned from the optical photometry of galaxies. Among early-type dwarf galaxies, we classify nucleated dwarf galaxies (dE,N and dS0, N) where an unresolved stellar nucleus is seen at the centre of the galaxy. Irregular galaxies are divided into HSB irregulars (HSB Irr) and LSB irregulars (LSB Irr) according to overall surface brightness of galaxies by visual inspection. Some irregular galaxies present elliptical shapes in outer part, but they show wide-spread irregular structures at their centres which are comparable to or larger than their half-light radius. We conducted a thorough classification based on their morphology and also deliberately avoided intermediate classes like `dE/Irr'. In Fig. 4, we present SDSS colour and $r-$band images of galaxies from the different morphological classes.

\subsection{The Ursa Major cluster catalogue }
The new catalogue of the Ursa Major cluster galaxies contains 166 objects. Table 2 presents the full catalogue including fundamental information such as membership, morphology, and optical/UV photometry. The following is a brief description of each column of Table 2:

 Column 1 : ID number. ID is marked with the RA 

 Column 2 : NGC, UGC, or other name of galaxy from \citet{Tul96}, \citet{Kar13}, and NED

 Columns 3 - 4 : right ascension (J2000) and declination (J2000) from SE{\tiny{XTRACTOR}} in degrees

 Columns 5 - 6 : supergalactic longitude and latitude in degrees

 Column 7 : heliocentric radial velocity (V$_h$) from the SDSS in km s$^{-1}$ (flag N : adopted from NED)

 Column 8 : radial velocity that adjusted for the frame of the Local Group by adding 300sin\textit{l}cos\textit{b} in km s$^{-1}$ (V$_{0}$)

 Column 9 : membership of galaxy from different catalogue (1 = this study, 2 = \citealt{Tul96}, and 3 = \citealt{Kar13}) 

 Column 10 : name of group defined by \citet{Kar13} which the galaxy belongs to

 Column 11 : morphology of galaxy classified from this study

 Column 12 : RC3 T-type from \citet{Tul96}

 Column 13 : RC3 T-type from \citet{Kar13}

 Columns 14 $-$ 18 : SDSS $u$, $g$, $r$, $i$, $z$ magnitudes and magnitude errors

 Columns 19 $-$ 20 : $GALEX$ NUV and FUV magnitudes and magnitude errors

 Column 21 : note for blue-cored early-type dwarf galaxies

\begin{landscape}
\begin{table}
\caption{The Ursa Major cluster catalogue}
\tiny
\begin{tabular}{@{}cccccclccccccrrrrrrrc}
\hline \hline
ID & Name & RA & Dec. & SGL & SGB & \multicolumn{1}{c} {V$_{h}$} & V$_{0}$ & Memb & Group & Morp & T & T & $u$ & $g$ & $r$ & $i$ & $z$ & NUV & FUV & Note \\
 &  & (deg) & (deg) & (deg) & (deg) & \multicolumn{1}{c}{(km s$^{-1}$)} & (km s$^{-1}$) & & K13 & & T96 & K13 & (mag) & (mag) & (mag) & (mag) & (mag) & (mag) & (mag) &  \\
(1) & (2) & (3) & (4) & (5) & (6) & \multicolumn{1}{c}{(7)} & (8) & (9) & (10) & (11) & (12) & (13) & (14) & (15) & (16) & (17) & (18) & (19) & (20) & (21) \\

\hline\hline
U001 & CGCG268-012 & 169.2511 & 50.5847 & 61.7831 & -2.5405 & $840$ &  905 & 1 & - & Irr(HSB) &         - &            9 & 16.26$\pm$0.03 & 15.23$\pm$0.02 & 14.83$\pm$0.02 & 14.69$\pm$0.02 & 14.43$\pm$0.03 & 17.32$\pm$0.04 & 17.75$\pm$0.12 & -\\
U002 & NGC3631 & 170.2620 & 53.1702 & 59.8040 & -0.7627 & $1155$ & 1233 & 1 & N3631 & Sc &         - &            5 & 12.10$\pm$0.03 & 10.96$\pm$0.02 & 10.53$\pm$0.02 & 10.27$\pm$0.02 & 10.18$\pm$0.03 & 12.87$\pm$0.03 & 13.26$\pm$0.08 & -\\
U003 & UGC6399 & 170.8469 & 50.8926 & 61.9877 & -1.5047 & $765$ &  833 & 1,2 & - & Sm &            9 &            8 & 15.01$\pm$0.03 & 13.99$\pm$0.02 & 13.56$\pm$0.02 & 13.32$\pm$0.02 & 13.17$\pm$0.03 & 16.02$\pm$0.04 & 16.30$\pm$0.09 & -\\
U004 & NGC3657 & 170.9817 & 52.9210 & 60.2259 & -0.4956 & $1208^N$ & 1286 & 1 & N3631 & Sb &         - &            0 & 14.45$\pm$0.04 & 13.15$\pm$0.02 & 12.59$\pm$0.02 & 12.30$\pm$0.02 & 12.12$\pm$0.03 & 15.24$\pm$0.04 & 15.37$\pm$0.09 & -\\
U005 & UGC6446 & 171.6685 & 53.7464 & 59.6786 &  0.2459 & $650$ &  733 & 1,2 & - & Sd &            7 &            7 & 14.22$\pm$0.03 & 13.35$\pm$0.02 & 13.21$\pm$0.02 & 12.90$\pm$0.02 & 13.25$\pm$0.03 & 14.50$\pm$0.03 & 14.71$\pm$0.07 & -\\
U006 & NGC3718 & 173.1459 & 53.0682 & 60.6766 &  0.7354 & $993^N$ & 1074 & 1,2,3 & N3992 & Sa &            1 &            1 & 13.18$\pm$0.03 & 11.23$\pm$0.02 & 10.51$\pm$0.02 & 10.08$\pm$0.02 & 9.84$\pm$0.03 & 14.15$\pm$0.03 & 14.92$\pm$0.07 & -\\
U007 & NGC3726 & 173.3374 & 47.0283 & 66.1767 & -1.7650 & $851$ &  904 & 1,2,3 & N3877 & SBc &            5 &            5 & 11.90$\pm$0.09 & 10.66$\pm$0.02 & 10.16$\pm$0.02 & 9.90$\pm$0.02 & 9.78$\pm$0.06 & 12.78$\pm$0.03 & 13.15$\pm$0.08 & -\\
U008 & NGC3729 & 173.4554 & 53.1259 & 60.7047 &  0.9279 & $1010^N$ & 1092 & 1,2,3 & N3992 & SBa &            2 &            1 & 13.33$\pm$0.03 & 11.89$\pm$0.02 & 11.25$\pm$0.02 & 10.87$\pm$0.02 & 10.69$\pm$0.03 & 14.98$\pm$0.03 & 15.58$\pm$0.08 & -\\
U009 & NGC3733 & 173.7588 & 54.8484 & 59.2234 &  1.8255 & $1185^N$ & 1275 & 1 & - & Sd &         - &            6 & 14.71$\pm$0.29 & 12.91$\pm$0.07 & 12.58$\pm$0.07 & 12.11$\pm$0.07 & 12.71$\pm$0.41 & 14.51$\pm$0.03 & 14.73$\pm$0.07 & -\\
U010 & NGC3769 & 174.4340 & 47.8925 & 65.7085 & -0.7270 & $705$ &  763 & 1,2,3 & N3769 & SBb &            3 &            3 & 13.52$\pm$0.03 & 12.30$\pm$0.02 & 11.77$\pm$0.02 & 11.47$\pm$0.02 & 11.26$\pm$0.03 & 14.62$\pm$0.03 & 15.04$\pm$0.07 & -\\

\hline
\end{tabular}
 \medskip
 {\bf Note.} Table 2 is presented in its entirety in the electronic edition of the Monthly Notices of the Royal Astronomical Society. 
 A portion is shown here for guidance regarding its form and content.
 
\end{table}
\end{landscape}

\subsection{Comparison with previous catalogues}
\begin{figure*}
\centering
\includegraphics[scale=0.8]{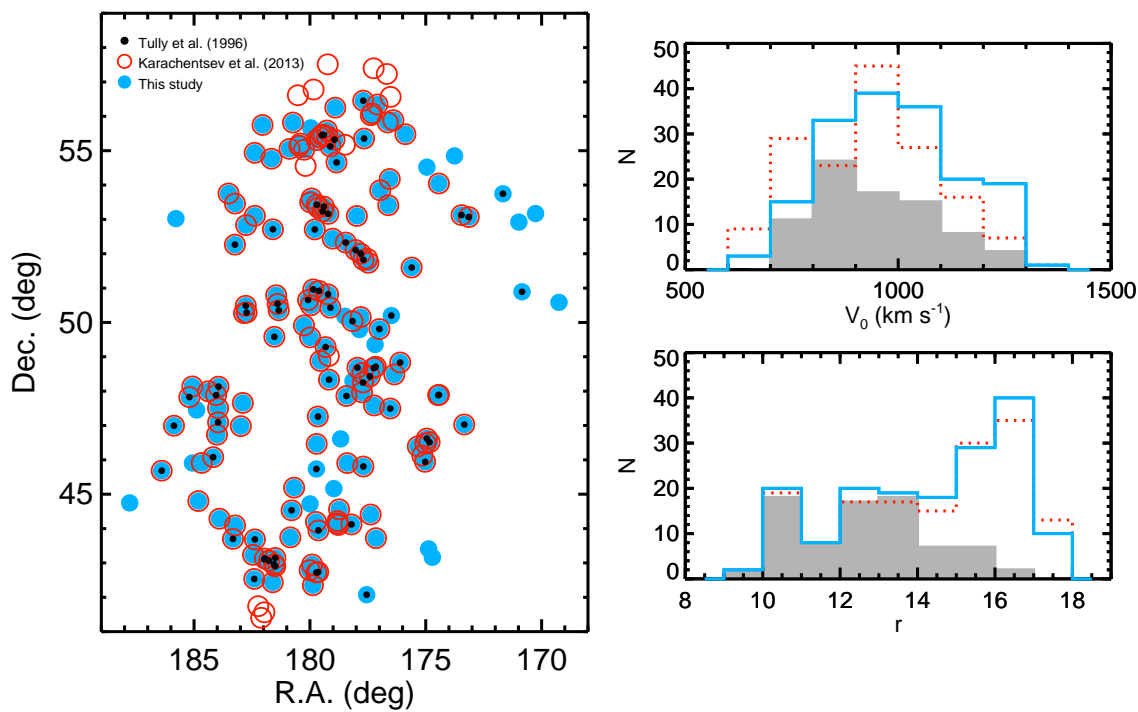}
\caption{Comparison of galaxies with those in previous catalogues. Left-hand panel: spatial distribution of galaxies. Blue filled circles are our galaxy sample, whereas black dots and red open circles are those from \citet{Tul96} and \citet{Kar13}, respectively. Right-hand panels: radial velocity distribution and absolute $r-$band luminosity function of galaxies in our catalogue (blue solid histogram ) in comparison with those of Tully et al. (1996, grey filled histogram) and Karachentsev et al. (2013, red dotted histogram).}\label{F5}
\end{figure*}

\begin{figure*}
\centering
\includegraphics[scale=0.7]{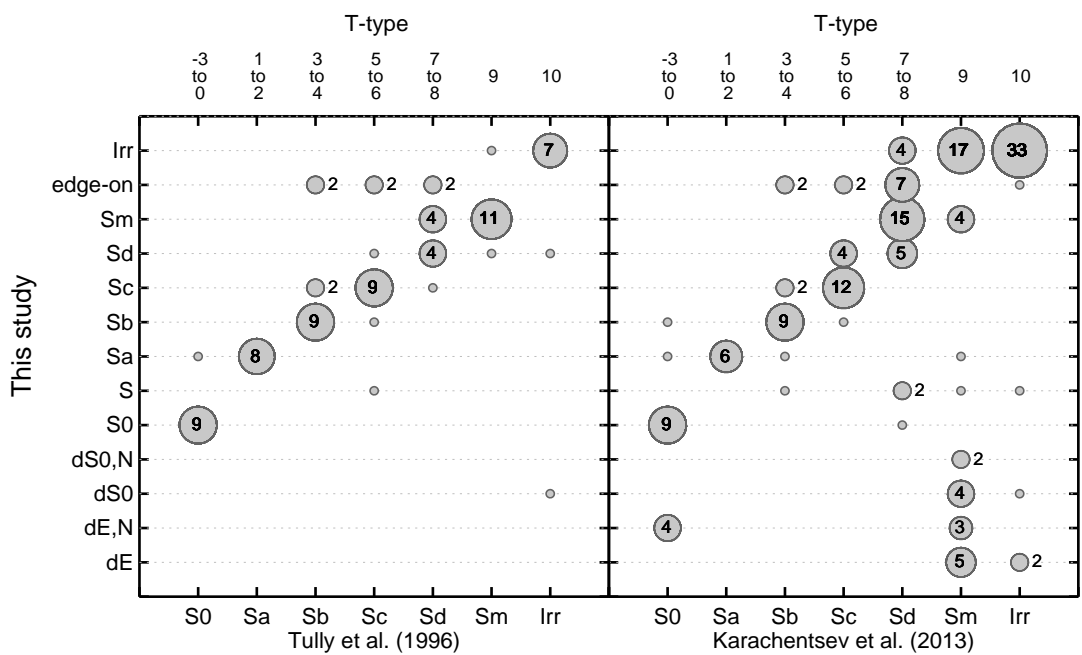}
\caption{Morphology comparison of this study with Tully et al. (1996, left) and Karachentsev et al. (2013, right). Point sizes correspond to the number of galaxies in each two-dimensional histogram cell. The number of galaxies in each cell is also indicated.}\label{F6}
\end{figure*}

\begin{figure*}
\centering
\includegraphics[scale=0.6]{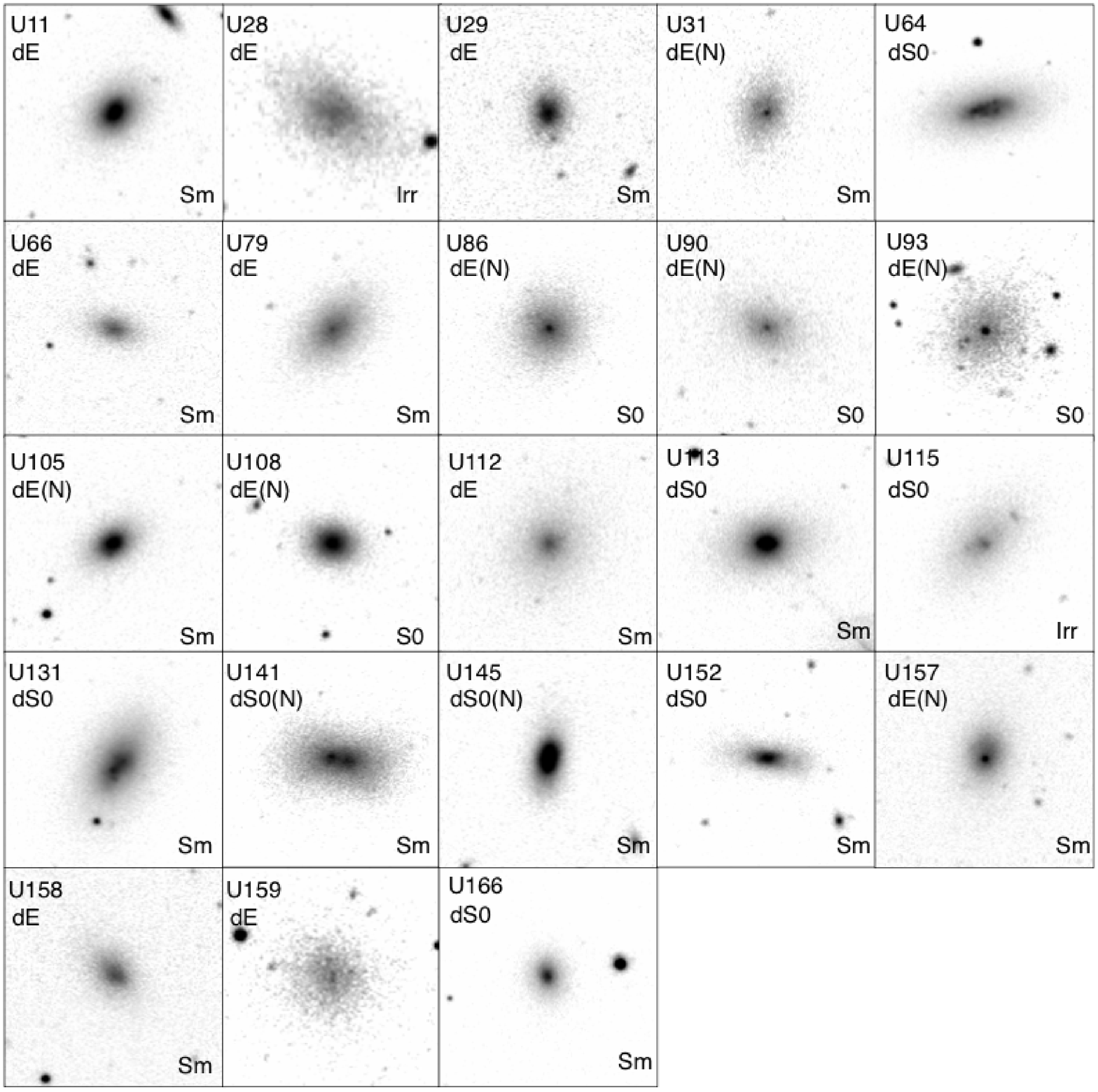}
\caption{SDSS $g-$, $r-$, $i-$combined images of 23 early-type dwarf galaxies in our catalogue. A large fraction of these galaxies are classified as different types in \citet{Kar13}. The ID of each galaxy is given in the upper left part of the image, along with our morphology. The morphology from \citet{Kar13}
is also given in the lower right part of the image.}\label{F7}
\end{figure*}

In Fig. \ref{F5} (left-hand panel), we present the spatial distribution of the 166 cluster galaxies (blue filled circles) included in our catalogue, in comparison with 79 and 157 member galaxies from Tully et al. (1996, black dots) and Karachentsev et al. (2013, open red circles), respectively. In the case of the catalogue of \citet{Kar13}, based on the galaxy group catalogue of \citet{Mak11}, they define the Ursa Major cluster consisting of seven galaxy subgroups: NGC 3769, NGC 3877, NGC 3992, NGC 4111, NGC 4157, NGC 4217, and NGC 4346 groups. From our 166 galaxies, 141 are also included in the seven groups of \citet{Kar13}. 

In the top right panel of Fig. \ref{F5}, we compare the velocity distribution of the galaxies between the three catalogues. At a glance, the strongly overlapping galaxy samples of all three catalogues cover a similar range in velocities. We obtain a velocity range and dispersion for the \citet{Kar13} sample as 653 $<$ V$_{0}$ $<$ 1361 km s$^{-1}$ and $\sigma_{V_{0}}$ = 158 km s$^{-1}$, which is in good agreement with those from ours and Tully et al. (1996, see also Sec. 2.1).
The bottom right panel of Fig. \ref{F5} shows a comparison of the r-band luminosity function among three catalogues. We adopt a distance modulus of the Ursa Major cluster of 31.20 from \citet{Tul12}. The luminosity functions of our sample and \citet{Kar13} are almost identical. Note that, since the galaxies selected from the SDSS spectroscopic data span a wide range in luminosity ($-21.5$ $<$ M$_r$ $<$ $-13.5$), the catalogues of ours and \citet{Kar13} contain a large fraction of faint dwarf galaxies (i.e., $-17$ $<$ M$_r$ $<$ $-13.5$) compared to  the sample of \citet{Tul96}. These faint galaxies mostly consist of early-type dwarf galaxies (dE and dS0) and Irr galaxies in our catalogue (see also Fig. \ref{F9}). 

In Fig. \ref{F6}, we present two-dimensional histograms comparing the morphologies of the galaxies in our catalogue with those of Tully et al. (1996, left-hand panel) and Karachentsev et al. (2013, right-hand panel). Adopting Table 1 in \citet{Nai10}, we translate the RC3 T-type number \citep{deV91} employed in the catalogues of \citet{Tul96} and \citet{Kar13} to the Hubble morphological classification scheme. The overall trend in Fig. \ref{F6} is that, except for dwarf galaxies, our morphological classification is largely consistent with those of \citet{Tul96} and \citet{Kar13}.

A significant disagreement occurs in the classes of Sm and Irr galaxies for \citet{Kar13}. A large fraction of these galaxies are classified as early-type dwarf galaxies (i.e., dE and dS0) in our catalogue (see lower-right part of right-hand panel of Fig. \ref{F6}). We point out that, while some of these galaxies show somewhat irregular features at their centres, all early-type dwarf galaxies show clear elliptical shapes with overall smooth and regular appearances.
We show in Fig. \ref{F7} the images of all these dwarf galaxies we classified as early-type. Tully \& Trentham\ (2008) noted an unusual population of central starburst dwarfs in the luminosity range of $-18$ $<$ M$_r$ $<$ $-14$ in the NGC5353/5354 group. These authors concluded that this population would be classified as dE based on their morphology. These galaxies look like some of our early-type dwarf galaxies with central irregularities, but are partly even bluer and more irregular. They may thus be considered transition-type dwarf galaxies.

\begin{figure}
\centering
\includegraphics[scale=0.4]{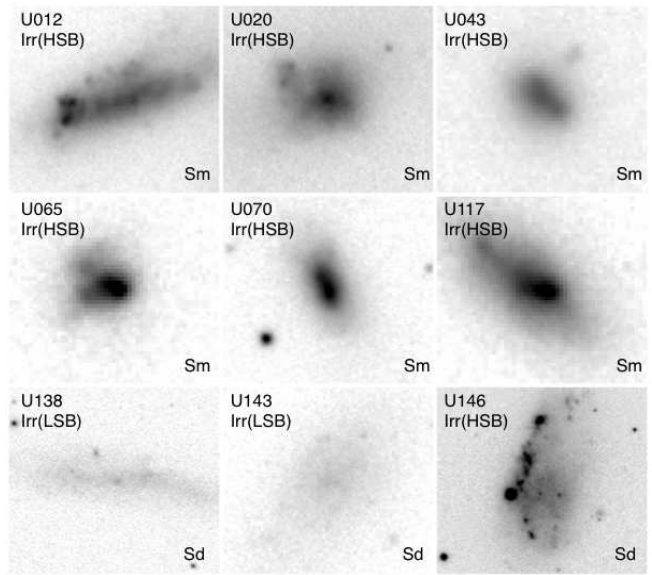}
\caption{SDSS $g-$, $r-$, $i-$combined images of irregular galaxies in our catalogue which are classified as different types in \citet{Kar13}. The ID of each galaxy is given in the upper left part of the image, along with our morphology. The morphology from \citet{Kar13}
is also given in the lower right part of the image. }\label{F8} 
\end{figure}

On the other hand, the classification comparison also shows a small scatter at fixed morphological type of our catalogue. A scatter is seen for edge-on galaxies in our catalogue, since we refrained from subdividing these further into subclasses. A difference with respect to \citet{Tul96} and \citet{Kar13} is also seen for our Irr galaxies in the sense that several of them have been classified as Sm or Sd in  \citet{Tul96} and \citet{Kar13} as shown in Fig. \ref{F8}.

\section{PROPERTIES OF THE URSA MAJOR CLUSTER}
\subsection{Type-specific luminosity function}
The type-specific luminosity function is helpful for studying galaxy evolution. Detailed studies of the luminosity function for different morphological types were performed for Virgo (\citealt{San85}; \citealt{Bin88}). The dedicated study by \citet{Jer97} found that the type-specific luminosity function of the Centaurus and Virgo cluster are very similar.

\begin{figure}
\centering
\includegraphics[scale=0.55]{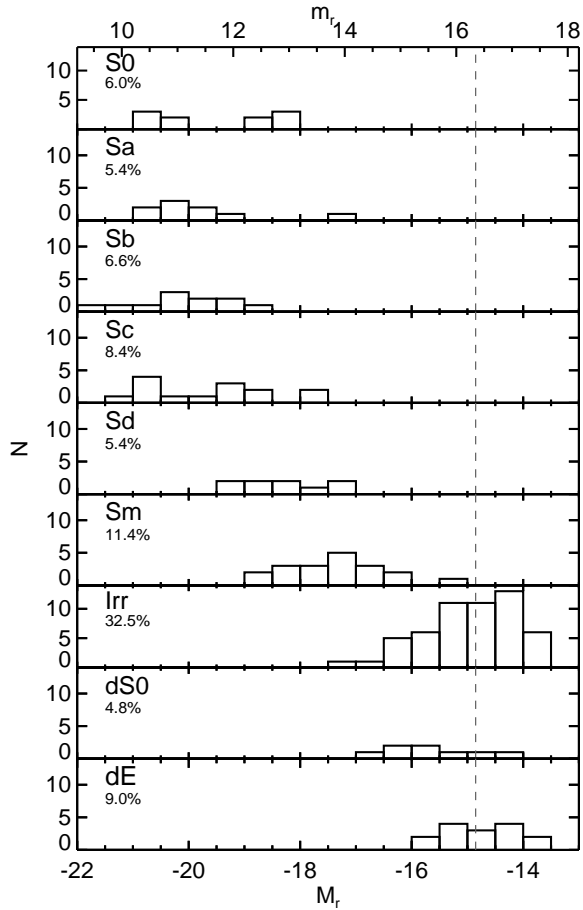}
\caption{Luminosity function of galaxies in our catalogue for various morphologies. In each panel, the fraction of each morphological type is indicated. The vertical dashed line is the magnitude (M$_r$ $= -14.85$) of the faintest galaxy in the catalogue of \citet{Tul96}}\label{F9}
\end{figure}

In Fig. \ref{F9}, we show the $r-$band type-specific luminosity function of the galaxies in the Ursa Major cluster. The Ursa Major cluster is dominated by late-type galaxies of spiral, edge-on, and irregular type (about 80 per cent, i.e., 133/166), but comprise a considerable fraction of early-type galaxies (about 20 per cent, i.e., 33/166). 

It is shown that the luminosity function depends on the galaxy type in the sense that Irr and late-type spiral galaxies are systematically fainter than S0 and early-type spiral galaxies (see also Sandage et al.\ 1985). In the case of early-type dwarf galaxies (dE and dS0), they are located in the  faintest range of the distribution. Irregular galaxies show a bell-shaped distribution and are skewed towards fainter magnitude. However, due to small number statistics, it is difficult to draw conclusions on the precise shape of the type-specific luminosity function for all galaxy types in the Ursa Major cluster.

\subsection{Spatial distribution} 
\begin{figure}
\centering
\includegraphics[scale=0.55]{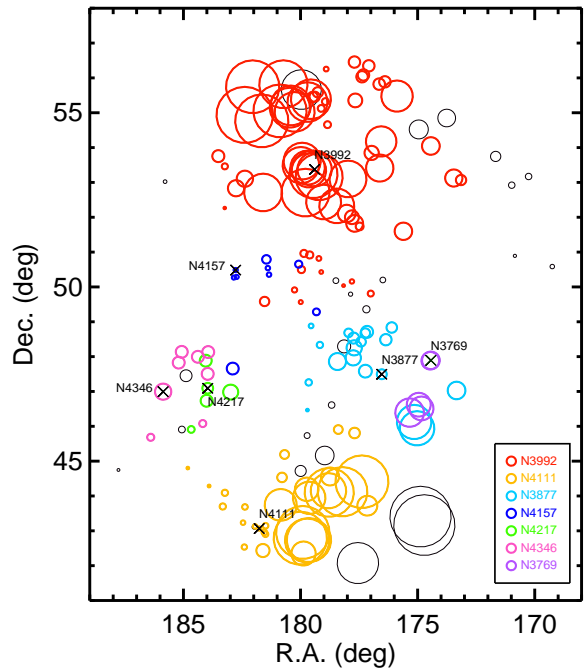}
\caption{Identification of subgroups in the Ursa Major cluster through the DS test bubble plot. The size of a circle is proportional to $\exp(\delta)$ (see the text). Different colours are for the seven subgroups adopted from \citet{Kar13}. The black circles represent galaxies which are not included in those seven subgroups. We also mark the name and location (cross) of the brightest galaxy of each subgroup.}\label{F10}
\end{figure}

\begin{figure}
\centering
\includegraphics[scale=0.8]{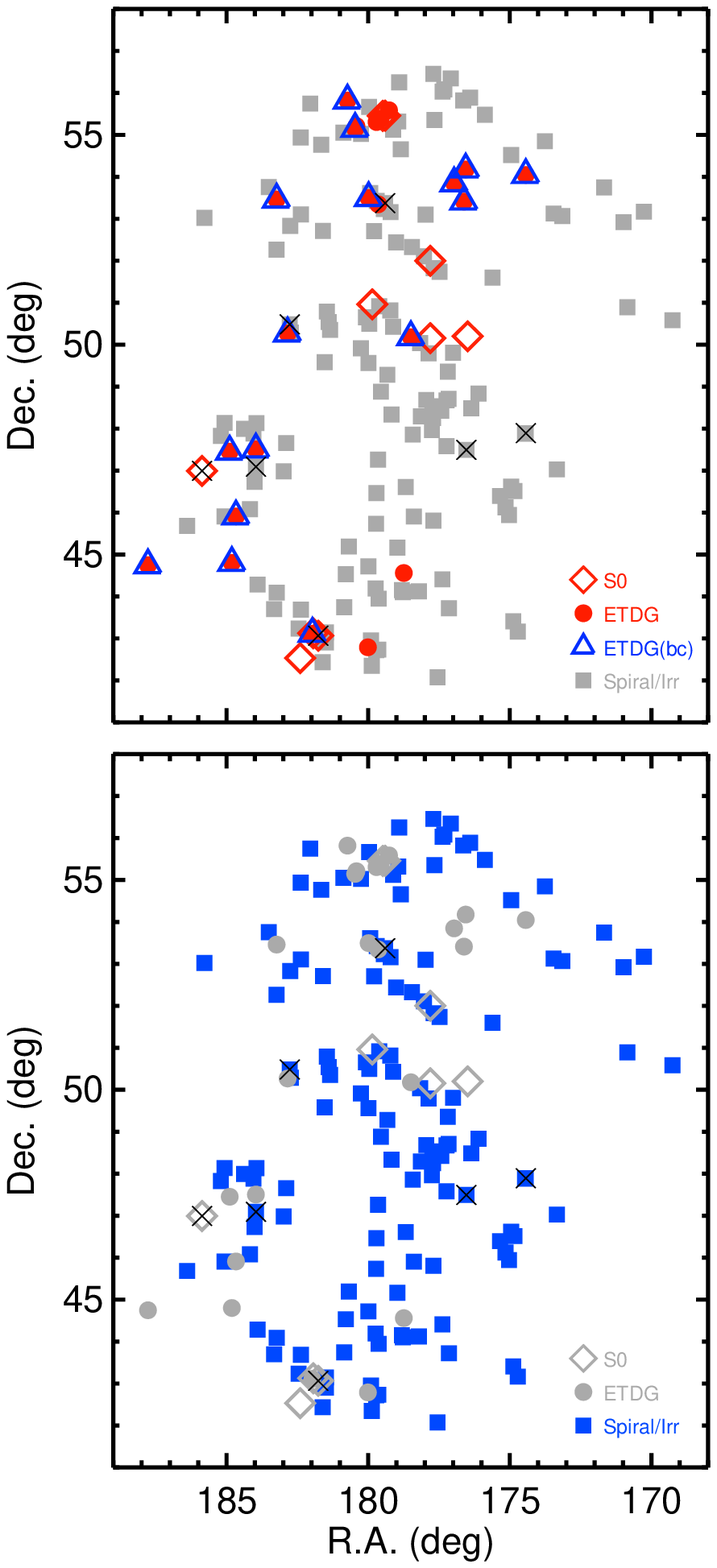}
\caption{ Projected spatial distribution of galaxies with different morphological types in the Ursa Major cluster. Both panels show the same galaxies but use different colour and symbols to highlight early-type galaxies (top) and late-type galaxies (bottom). In the legend given in the top panel, `ETDG' stands for early-type dwarf galaxy. In the top panel, ETDGs with blue cores (ETDG(bc)s) are additionally marked by triangles. The crosses indicate the brightest galaxy of each subgroup from \citet{Kar13}.}\label{F11}
\end{figure}

The fact that the overall spatial distribution of galaxies in the Ursa Major cluster is patchy (see Fig. \ref{F5}) is already a hint towards the existence of several subgroups. Based on the Local Supercluster galaxy group catalogue of \citet{Mak11}, which was constructed by an algorithm of the modified friend-of-friend method, \citet{Kar13} have suggested that the Ursa Major cluster includes seven subgroups. In order to confirm the existence of subgroups, we carried out the Dressler$-$Shectman test (DS test; \citealt{Dre88}) which allows us to quantify the substructure in a galaxy cluster using the line-of-sight velocities and the projected distribution of galaxies. The deviations with respect to the systemic velocity ($\textit{v}_{c}$) and the velocity dispersion ($\sigma_{c}$) of the cluster are calculated for each galaxy as

\begin{displaymath}  \delta^2 = \left( \frac{N+1}{ \sigma_{c}^{2} } \right) [(\textit{v}_{local} - \textit{v}_{c})^2 + (\sigma_{local} - \sigma_{c})^2],\end{displaymath}

where N is the number of the nearest galaxies, $\textit{v}_{local}$ and $\sigma_{local}$ are their mean velocity and velocity dispersion, respectively. Here we adopt N = $\sqrt{N_{total}}$ from \citet{Pin96}. N$_{total}$ is the total number of members in the cluster. In Fig. \ref{F10}, we present the DS test bubble plot for galaxies. The circle size is proportional to $\exp(\delta)$. This $\delta$ statistic is used as the significance of the substructure, measuring the local deviation of each galaxy from a smooth, virialized velocity and spatial distribution.

For comparison, we also overplot the seven subgroups of \citet{Kar13} with different colours. The location of the brightest main galaxy of each subgroup is also indicated with a cross. In the most prominent subgroup NGC 3992, the galaxies appear to be concentrated around the brightest galaxy of the subgroup. The central region of the Ursa Major cluster consists of galaxies showing small $\delta$ values indicated by small circles and is dominated by galaxies associated with NGC 3877 as well as with the NGC 4346/NGC 4217 subgroups. In the northern part of the cluster, there is a distinct subgroup with large $\delta$ values (red circles), which corresponds to the NGC 3992 subgroup. In addition, one subgroup is found in the southern region, which might be associated with NGC 4111 (yellow circles). While the substructure of a cluster can be barely distinguished by the velocity distribution alone, by means of DS test we conclude that the Ursa Major cluster does consist of several subgroups, confirming the findings of \citet{Kar13}. 

In Fig. \ref{F11}, we present the projected spatial distribution of early-type (S0, dE, and dS0; upper panel) and late-type (spiral and Irr; bottom panel) galaxies separately. As \citet{Kar13} noted, the galaxy distribution in the Ursa Major cluster is significantly elongated along the supergalactic equator (see also Fig. \ref{F2}), suggesting an unrelaxed, dynamically young structure. Late-type galaxies in the Ursa Major cluster are distributed over the whole area of the cluster. On the other hand, the early-type dwarf galaxies tend to be more clustered. The largest number of early-type dwarf galaxies are found towards the NGC 3992 subgroup. A two-dimensional Kolmogorov-Smirnov (KS) test on the spatial distributions between early-type galaxies and late-type galaxies yields a probability of only 2 per cent for the same parent distribution of the two subsamples.

\subsection{CMR}
\begin{figure}
\centering
\includegraphics[scale=0.6]{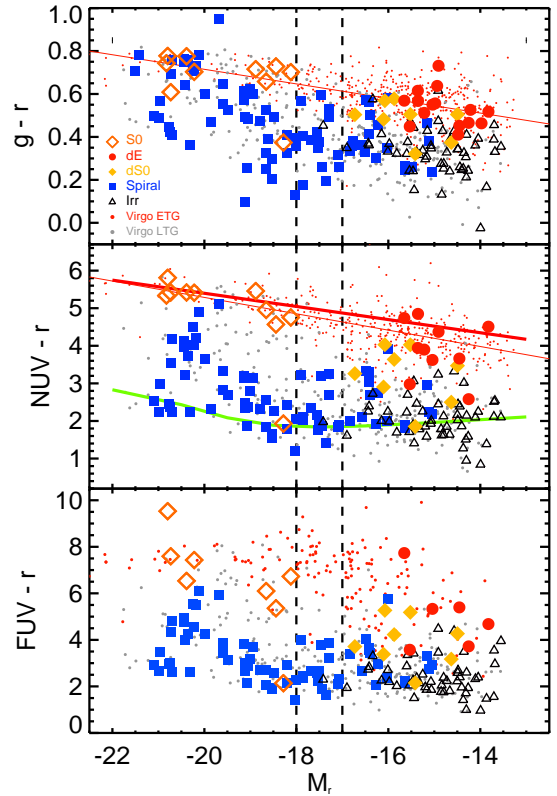}
\caption{Optical (top) and UV-optical (middle and bottom) CMRs of the Ursa Major cluster. Differently coloured symbols are galaxies with different morphologies in the Ursa Major cluster; grey and red dots are Virgo galaxies. In the legend given in the top panel, `Virgo ETG' and `Virgo LTG' stands for early and late-type galaxies in the Virgo cluster, respectively. In the $g-r$ and NUV$-r$ CMRs, the red thin solid lines represent the linear least-square fit to the early-type galaxies in the Virgo cluster. In the NUV$-r$ CMR, the red and green thick solid lines indicate the red sequence and blue cloud defined by \citet{Wyd07}. The two vertical dashed lines indicate the magnitude range ($-18$ $<$ M$_r$ $<$ $-17$) of a possible gap between S0s and early-type dwarf galaxies in the Ursa Major cluster. RMS error bars for Ursa Major bright (M$_r$ $<$ $-18$) and faint (M$_r$ $>$ $-18$) galaxies are shown in the upper left and right corner of each CMR panel. Note that error bars in the UV CMRs are omitted because their sizes are much smaller than the symbol size. Even for faint red galaxies, colour errors rarely exceed 0.1 and 0.2 mag for NUV$-r$ and FUV$-r$, respectively. }\label{F12}
\end{figure}

In Fig. \ref{F12}, we present $g-r$, NUV$-r$, and FUV$-r$ colour-magnitude relations (CMRs) of galaxies in the Ursa Major cluster (large coloured symbols). For comparison with galaxies in a different environment, we also overplot CMRs of  the Virgo cluster (dots) based on SDSS optical and $GALEX$ UV photometry of galaxies in the Virgo cluster catalogue (see \citealt{Kim10}). We adopt a distance modulus of the Virgo cluster of 31.01 from \citet{Tul12}.

The CMRs of galaxies in the Ursa Major cluster follow the general trend in the sense that galaxies become progressively bluer with decreasing optical luminosity. At all luminosities, the Ursa Major cluster sample is dominated by late-type galaxies in the blue cloud of the CMRs, in contrast to the Virgo cluster. In the optical CMR, the early-type galaxies (S0/dE/dS0) of the Ursa Major cluster exhibit a well-defined sequence. While they do follow closely the red sequence of their Virgo counterparts (red thin solid line in Fig. \ref{F12}), the early-type dwarf galaxies with M$_r$ $>$ $-17$ mag show a small blueward offset: 57 per cent lie blueward of the Virgo red sequence in g-r (top panel). This offset becomes more prominent when using UV-optical colours: 78 per cent of the Ursa Major early-type dwarf galaxies lie blueward of the Virgo red sequence in NUV$-r$ (middle panel). They occupy the region between red sequence and blue cloud as defined by Wyder et al. (2007, red and green thick solid lines), which was based on galaxies in the local Universe with a redshift range of 0.01 $<$ z $<$ 0.25 and $r-$band magnitudes in the range of 14.5 $<$ r $<$ 17.6.

Note that some fraction of the faint early-type dwarf galaxies in the Virgo cluster also show blue NUV$-r$ and FUV$-r$ colours. It has been suggested that most of these galaxies are in the outer region ($>$ 2$\degr$ from M87) of the Virgo cluster and follow slightly steeper NUV$-r$ and FUV$-r$ CMRs than those of their counterparts in the inner region (see \citealt{Kim10} for details). In the present case, we find that the faint early-type dwarf galaxies in the Ursa Major cluster have bluer UV$-r$ colours than counterparts in the outskirts of the Virgo cluster. We calculate mean values of NUV$-r$ colour of early-type dwarf galaxies in the Ursa Major and Virgo outskirts are 3.7 and 4.2 with standard deviations of 0.81 and 0.63 mag, respectively, in the magnitude range of $-17$ $<$ M$_{r}$ $<$ $-13$.

The UV CMRs are particularly efficient tools to trace recent star formation, owing to their sensitivity to young ($<$ 1 Gyr) stellar populations (e.g., \citealt{Yi05}; \citealt{Kav07}). This indicates that most faint early-type dwarf galaxies in the Ursa Major cluster have undergone recent star formation (\citealt{Bos05}; \citealt{Lis06}; \citealt{Kim10}; \citealt{Hal12}, see also Sec. 4.4).

\begin{figure*}
\centering
\includegraphics[scale=0.8]{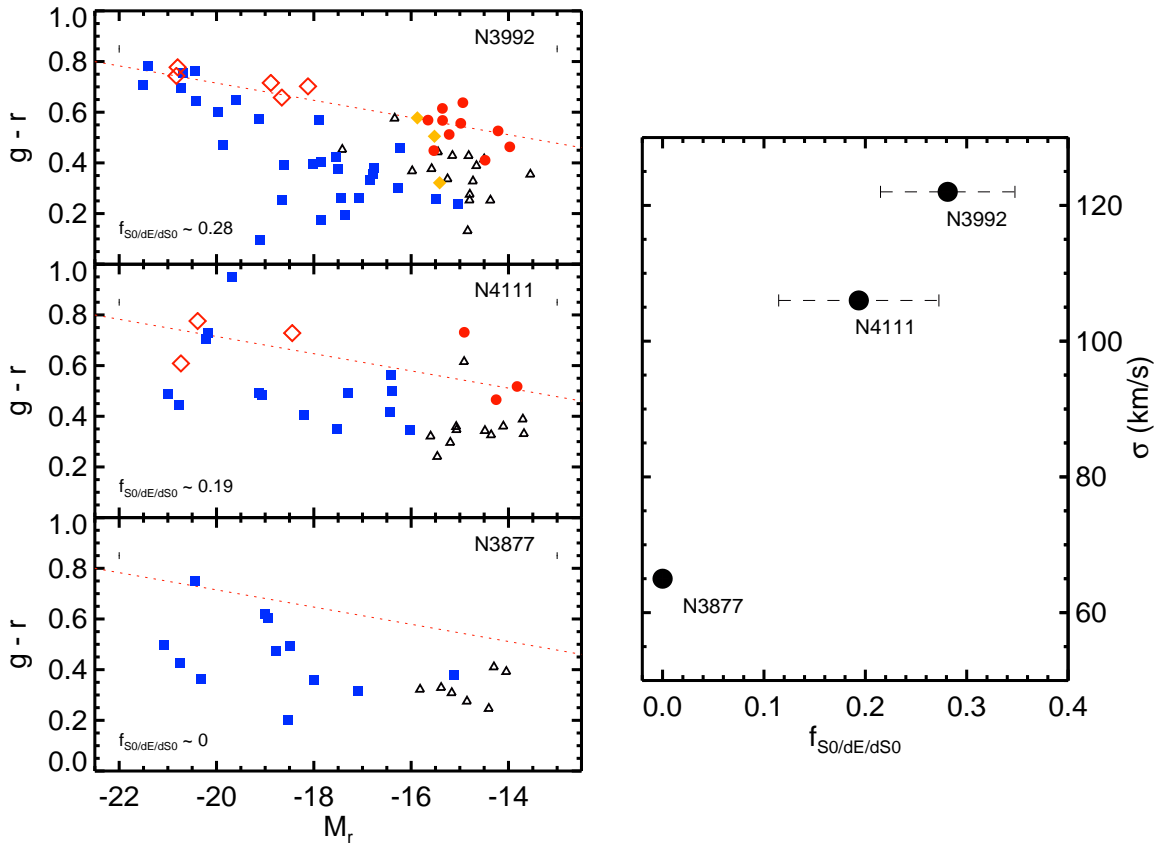}
\caption{Left: optical colour$-$magnitude diagrams for the three most massive subgroups in the Ursa Major cluster. Symbols and error bars are the same as in Fig. \ref{F12}. The dotted line is the red sequence of the Virgo cluster as a reference. In each panel, the fraction of early-type galaxies (f$_{S0/dE/dS0}$) is given in the lower left corner. rms error bars for Ursa Major bright (M$_r$ $<$ $-18$) and faint (M$_r$ $>$ $-18$) galaxies are shown in the upper-left and upper-right corner of each CMR panel. Right: the fraction of early-type galaxies (filled circles) against the velocity dispersion of the subgroup. The Poisson error for each fraction is also given.}\label{F13}
\end{figure*}

We present the $g-r$ CMR separately for the three most massive subgroups in the Ursa Major cluster (NGC 3992, NGC 4111, and NGC 3877; see left-hand panels of Fig. \ref{F13}). Due to small number statistics, the other four  subgroups are not considered. We adopt the velocity dispersions of the subgroups reported in \citet{Kar13}. We determine the mean fraction of early-type galaxies, f$_{S0/dE/dS0}$, for the three subgroups. The NGC 3992 group, the most massive subgroup with $\sigma_{V}$ = 122 km s$^{-1}$, has a relatively well populated red sequence with f$_{S0/dE/dS0}$ $\sim$ 0.28 (i.e., 18/64). About half (55 per cent, 18/33) of the early-type galaxies in the Ursa Major cluster seem to be associated with the NGC 3992 group. The second massive subgroup, NGC 4111 ($\sigma_{V}$ = 106 km s$^{-1}$), has f$_{S0/dE/dS0}$ $\sim$ 0.19 (i.e., 6/31), whereas NGC 3877 ($\sigma_{V}$ = 65 km s$^{-1}$) contains only late-type  galaxies. This indicates a correlation between the fraction of early-type dwarf galaxies with the velocity dispersion of the subgroup (see right-hand panel of Fig. \ref{F13}). Our result is in accordance with previous results based on the massive early-type sample in poor groups (e.g., \citealt{Tov04}; \citealt{Tov09}).

\subsection{Blue-cored early-type dwarf galaxies}
\begin{figure*}
\centering
\includegraphics[scale=0.5]{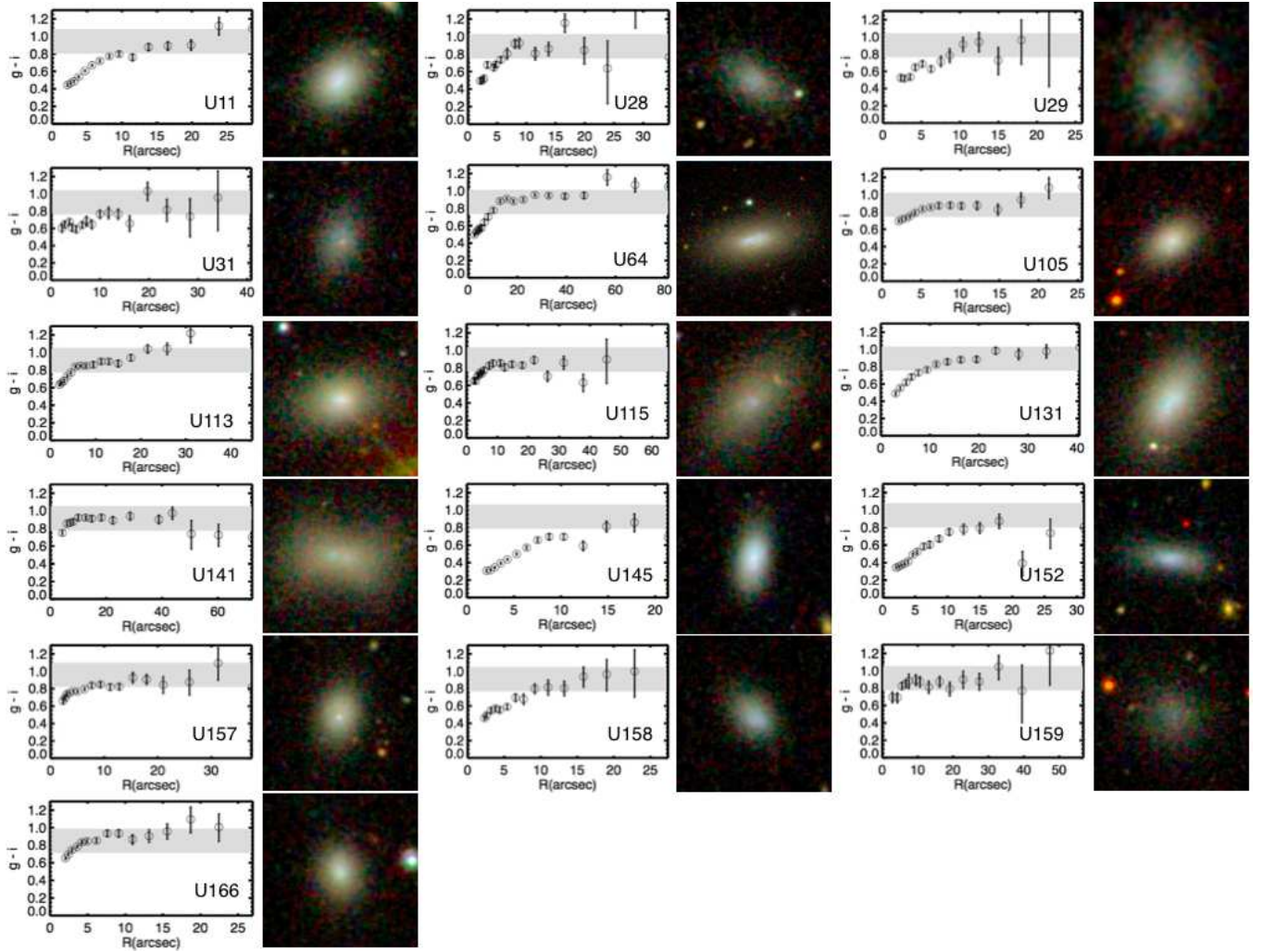}
\caption{Radial $g-i$ colour profiles and images of the 16 blue-cored early-type dwarf galaxies in our catalogue. In the colour profile, the radius is calculated from semi-major (a) and semi-minor (b) axes as (ab)$^{1/2}$. The error bars denote the uncertainty calculated from magnitude errors. The grey-shaded areas include the 2$\sigma$ range of the colours of normal dEs without blue core at the respective magnitude derived from the Virgo cluster CMR of \citet{Lis08}.}\label{F14}
\end{figure*}

\begin{figure*}
\centering
\includegraphics[scale=0.4]{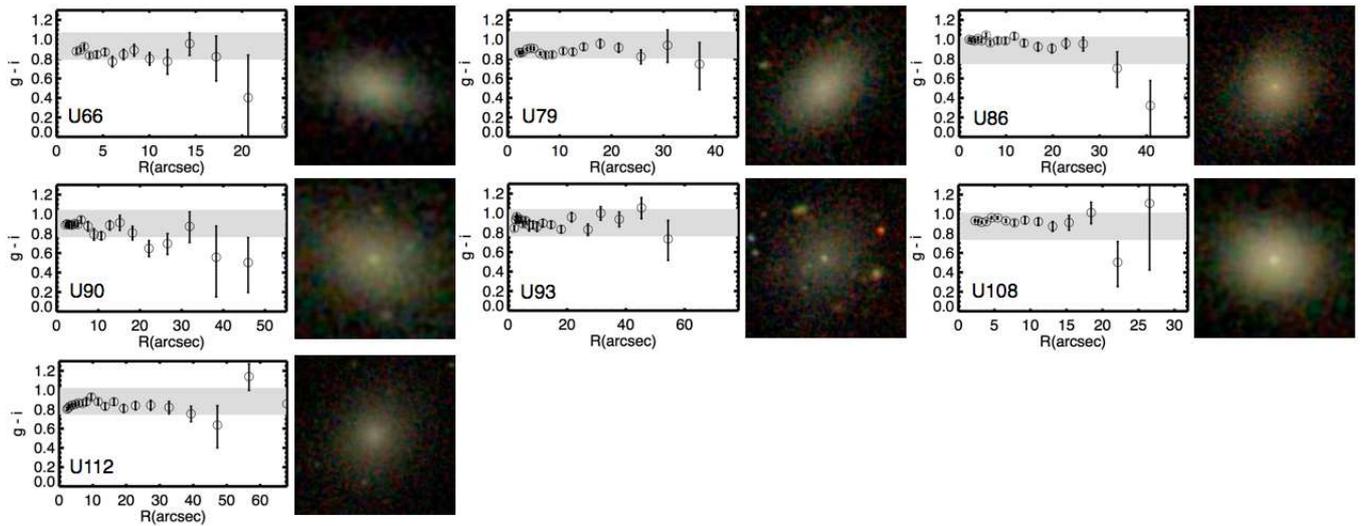}
\caption{Same as Fig. \ref{F13}, but for the seven early-type dwarf galaxies without blue core.}\label{F15}
\end{figure*}

\begin{figure}
\centering
\includegraphics[scale=0.8]{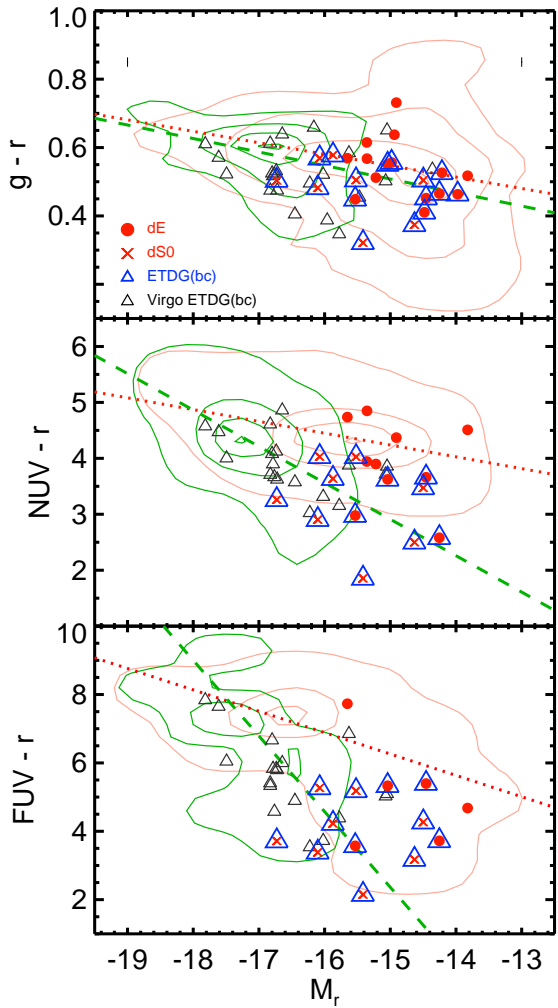}
\caption{Optical and UV colour$-$magnitude relations of early-type dwarfs in the Ursa Major cluster. The dEs and dS0s are represented by red filled circles and red crosses, respectively. The blue-cored early-type dwarf galaxies (ETDG(bc)) are additionally marked by blue triangles. For comparison, the distributions of dEs and dS0s in the Virgo cluster are overplotted by red and green contours, respectively. The red dotted line and green dashed line represent linear least-squares fits to the dEs and dS0s in the Virgo cluster, respectively (Kim et al. 2010). The black open triangles indicate blue-cored early-type dwarfs in the Virgo cluster defined by \citet{Lis06}. rms error bars for Ursa Major bright (M$_r$ $<$ $-19$) and faint (M$_r$ $>$ $-19$) galaxies are shown as error bars in the upper-left and upper-right corner of each CMR panel.}\label{F16}
\end{figure}

\begin{figure}
\centering
\includegraphics[scale=0.4]{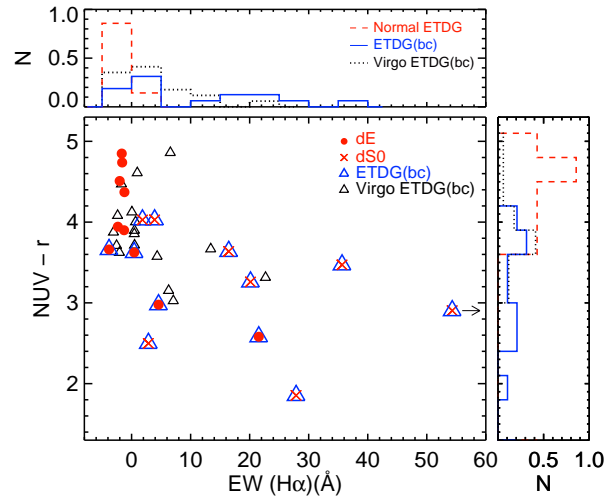}
\caption{NUV$-r$ colour versus EW (H$\alpha$) for early-type dwarfs in the Ursa Major cluster, in comparison to Virgo blue-cored early-type dwarfs (black triangles). The EW (H$\alpha$) of all galaxies are extracted from the SDSS archive. The distributions of NUV$-r$ and EW (H$\alpha$) of galaxies are presented in the right and upper histograms, respectively.}\label{F17}
\end{figure}

\begin{figure}
\centering
\includegraphics[scale=0.4]{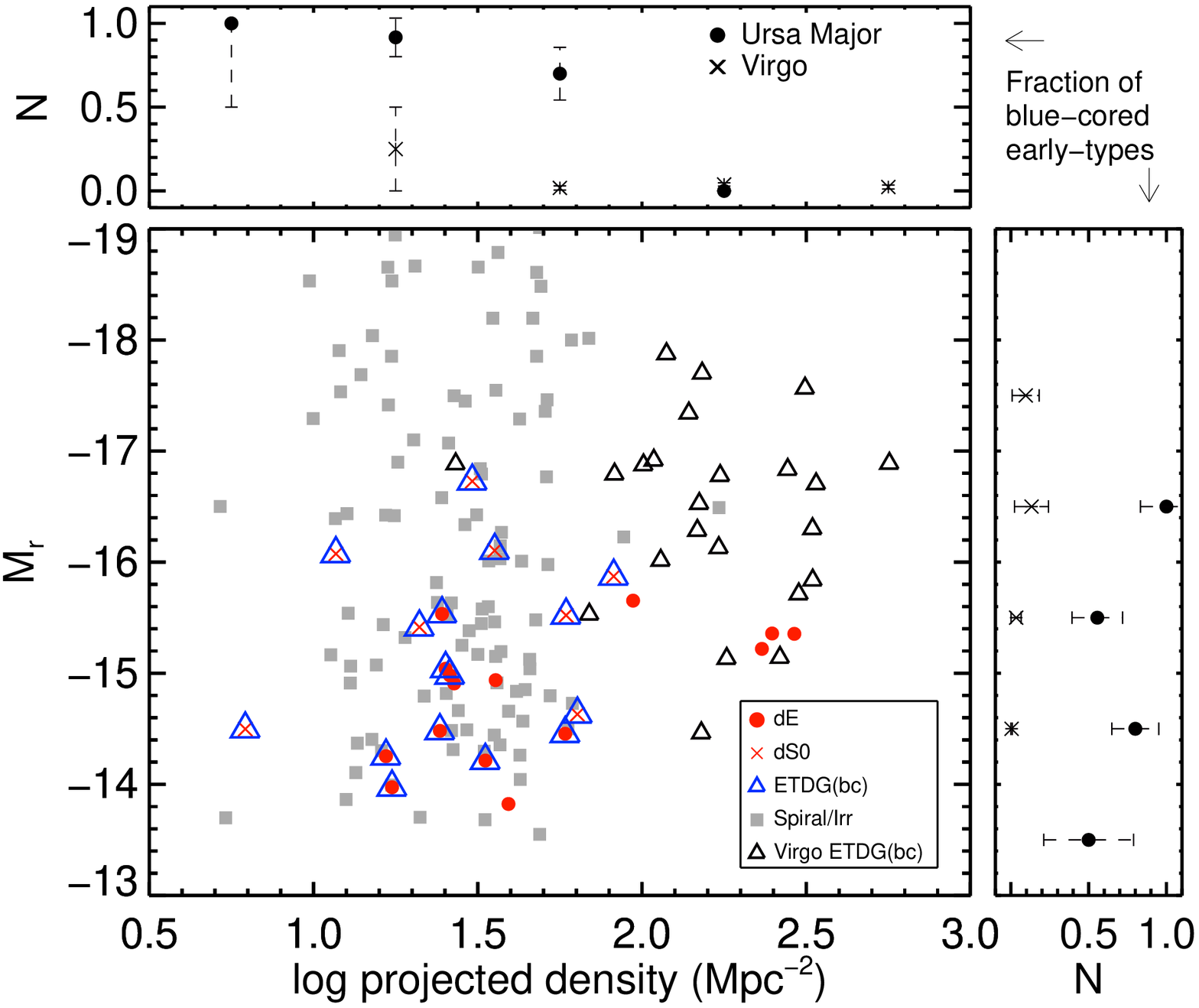}
\caption{Local projected galaxy density versus M$_r$ for early-type dwarf galaxies in the Ursa Major cluster. The blue-cored early-type dwarfs and late-type galaxies in the Ursa Major cluster are denoted by blue triangles and grey filled boxes, respectively. For comparison, we also plot blue-cored early-type dwarfs in the Virgo cluster (black triangles). The adjacent panels show blue-cored early-type fraction of the Ursa Major cluster and Virgo cluster as a function of density (upper) and M$_r$ (right), using Poisson errors for the Virgo cluster fractions and binomial errors for the Ursa Major cluster fractions.}\label{F18}
\end{figure}

In order to confirm the presence of blue cores and positive colour gradients, we have examined the radial $g-i$ colour profiles of all Ursa Major early-type dwarf galaxies by performing surface photometry. For that purpose, background subtracted $g-$, $r-$, $i-$band images were aligned and stacked to increase the S/N ratio. Elliptical annuli for each galaxy were determined using the ellipse fitting procedure of {\tiny{IRAF}} ELLIPSE \citep{Jed87}.

In the ellipse fit, we fixed the position angle, ellipticity, and centre of the isophotes using the values returned from SE{\tiny{XTRACTOR}}. We then measured separately the $g$ and $i-$band surface brightness profiles using the same elliptical annuli, so that the $g-i$ colour is derived from the same region.

Out of 23 early-type dwarf galaxies, 16 galaxies show a blue cores based on their radial $g-i$ colour profiles (see Fig. \ref{F14}). While some of those 16 galaxies have relatively weak colour gradients (e.g., U31 and U159), most galaxies with blue cores exhibit colour differences of more than 0.2 mag between the central and outer regions. In several galaxies, central irregularities are also identified, which are presumably caused by gas and dust associated with ongoing or recent star formation activity (e.g., \citealt{DeL10} for Virgo blue-cored early-type dwarfs). In Fig. \ref{F15} we also present the $g-i$ colour profiles and images of the seven early-type dwarf galaxies without blue cores.

In Fig. \ref{F16}, we present $g-r$, NUV$-r$, and FUV$-r$ CMRs for the dEs (red circles) and dS0s (red crosses) in the Ursa Major cluster. For comparison, we overplot the distribution of the dEs (red contours) and dS0s (green contours) in the Virgo cluster. The Virgo dEs and dS0s show a similar distribution in the $g-r$ CMRs (upper panel). However, the dS0s form a distinct sequence in the UV-optical CMRs (middle and bottom panels) which has a steeper slope compared to the dEs (see also \citealt{Kim10}). The dS0s in the Ursa Major cluster follow the dS0 sequence of the Virgo cluster at the faint end in the UV-optical CMRs. Overall, dS0s in the Ursa Major cluster show bluer UV-optical colours than the dEs at a given magnitude, which is also found in the Virgo cluster. On the other hand, a large fraction (83 per cent) of Ursa Major dEs are bluer than the mean distribution of Virgo dEs, especially in FUV$-r$ CMR.
  
Since the UV flux is particularly sensitive to the presence of young ($<$1 Gyr) stars, the blue UV-optical colours of blue-cored early-type dwarfs indicate that these galaxies have experienced recent or ongoing star formation activities in their central regions. In order to confirm this, we examined integrated SDSS spectra. The SDSS spectroscopic fibre diameter (3 arcsec) corresponds to 253 pc at the distance of the Ursa Major cluster, providing spectral information for the very central region of the galaxy. In Fig. \ref{F17}, we show the distribution of equivalent width of the H$\alpha$ emission line (EW (H$\alpha$)) for early-type dwarfs against the NUV$-r$ colour. H$\alpha$ emission traces  the most massive O- and B-type stars of ages less than a few million years \citep{Ken98}.

As expected, NUV$-r$ colour is correlated with EW (H$\alpha$) in the sense that bluer galaxies show higher EW (H$\alpha$). The majority of blue-cored early-type dwarfs (triangles) have a different distribution of EW (H$\alpha$) values than those of early-type dwarfs without blue core (circles and crosses without triangles). We note that the blue-cored early-type dwarfs tend to show more enhanced H$\alpha$ emission line strengths (see also upper histograms). We find that 75 per cent of NUV-detected blue-cored early-type dwarfs exhibit residual star formation with EW (H$\alpha$) $>$ 2 \AA, which is a lower limit for star forming activity following \citet{Bos08}. Furthermore, blue-cored early-type dwarfs in the Ursa Major cluster are distributed towards bluer NUV$-r$ colour and higher EW (H$\alpha$) values than their Virgo counterparts, indicating more active recent or ongoing star formation in the blue-cored early-type dwarfs in the Ursa Major cluster (see right histograms in Fig. \ref{F17}).

In Fig. \ref{F18} we present the projected local densities against the luminosities of the early-type dwarfs in the Ursa Major cluster comparing with blue-cored early-type dwarfs in the Virgo cluster. We also compare the densities of blue-cored early-type dwarfs with the distribution of spiral and irregular galaxies in the Ursa Major cluster. Following \citet{Dre80}, we calculate the projected local density of each galaxy by defining a circular area around  the galaxy that encloses the 10th nearest-neighbour galaxy. The projected local density is then the number of galaxies per square Mpc within this circle. First of all, the density distribution of blue-cored early-type dwarfs in the Ursa Major cluster shows no significant difference with that of spiral and irregular galaxies. 
A KS test gives the probability of 27 per cent that the densities of blue-cored early-type dwarfs and spiral/irregular galaxies are drawn from the same distribution. This suggests that blue-cored early-type dwarfs share similar density environments of late-type galaxies.

\citet{Lis06, Lis07} reported that the blue-cored early-type dwarfs in the Virgo cluster are found in moderate to low-density cluster regions with no centrally clustered projected spatial distribution, whereas normal early-type dwarfs dominate the high-density regions. Note that blue-cored early-type dwarfs in the Ursa Major cluster systematically reside in the lower density environments (about a factor of 6) compared to those in the Virgo cluster. This indicates that these galaxies are also frequently found in more sparsely populated  regions compared to  the centre of the cluster.
 On the other hand, the blue-cored early-type dwarfs in the Ursa Major cluster are distributed towards lower luminosity compared to those in the Virgo cluster. Note that Virgo blue-cored early-type dwarfs are restricted to brighter magnitudes with m$_{B}$ $<$ 16, and the decline of the number fraction of these galaxies as a function of magnitude is real and not caused by S/N effects (see \citealt{Lis06} for details).

\section{DISCUSSION AND CONCLUSION}
  We present the results of optical and ultraviolet photometric properties of the Ursa Major cluster galaxies. Also we performed a comparison study of the Ursa Major cluster and the Virgo cluster to investigate the environmental dependence of galaxy properties.

\begin{enumerate}[I.]
\item We have constructed a new catalogue of the Ursa Major cluster galaxies with available spectroscopic data from SDSS and NED. Membership was defined within a 7$\degr$.5 radius circle centred at RA(B1950) = 11$^h$ 56$^m$.9 and Dec.(B1950) = $+$49$\degr$ 22$\arcmin$, following \citet{Tul96}, and a velocity range of 677 $<$ V$_{0}$ $<$ 1302 km s$^{-1}$. In total 166 galaxies brighter than M$_{r} = -13.5$ mag were selected as members, more than twice as many as reported in \citet{Tul96}. Out of our 166 galaxies, 77 and 141 ones are also included as members in \citet{Tul96} and \citet{Kar13}, respectively. 

\item We derived SDSS optical and $GALEX$ UV photometric parameters of all cluster members. Morphological classification was carried out by examining SDSS images. Except for early-type dwarf galaxies, our morphological classification is largely consistent with those of \citet{Tul96} and \citet{Kar13}. Our new catalogue of the Ursa Major cluster includes fundamental information on galaxies such as morphology, radial velocity, and optical/UV photometry.

\item We found evidence for the existence of several subgroups in the Ursa Major cluster based on the DS test, confirming previous findings of \citet{Kar13}. While the Ursa Major cluster is dominated by late-type galaxies, in the most massive subgroup NGC 3992 has a relatively large population of early-type galaxies (S0/dE/dS0) with early-type fraction of ~ 28 per cent (18/64). About half of all early-type galaxies (55 per cent, 18/33) in the Ursa Major cluster are members of this subgroup. The second massive subgroup around NGC 4111 has an early-type fraction of $\sim$19 per cent (6/31). The NGC 3769 and NGC 3877 subgroups have only late-type galaxies. 

\item We analysed the CMRs for galaxies in the Ursa Major cluster and compared them with the Virgo cluster, using the $g-r$, NUV$-r$, and FUV$-r$ colours as a function of absolute $r-$band magnitude. The Ursa Major cluster is significantly underpopulated in the red sequence. In particular, there are no luminous early-type dwarf galaxies (brighter than M$_{r} \sim$ $-17$ mag) in the cluster. In the $g-r$ CMR, the slopes of the red sequences of the two cluster populations are very similar. The faint (M$_{r}$ $>$ $-17$) early-type dwarf galaxies lie blueward of the Virgo red sequence in UV$-r$ CMRs.

\item We discover blue-cored early-type dwarf galaxies, which are already known to exist in the outskirts of the Fornax and Virgo clusters (\citealt{DeR13}; \citealt{Lis06}) as well as in the field \citep{Gu06}. Out of 23 Ursa Major early-type dwarf galaxies, 16 galaxies ($\sim$70 per cent) show evidence of central star formation. The fraction of blue-cored early-type dwarfs in the Ursa Major cluster is much higher than that of the Virgo cluster and extends to fainter magnitude. Their overall UV-optical colours are consistent with the dS0 sequence of the Virgo cluster at the faint end.

\end{enumerate}

\subsection{Presence of blue-cored early-type dwarf galaxies in the Ursa Major cluster}
The most striking result of our study is that about 70 per cent of the early-type dwarf galaxy population in the Ursa Major cluster have blue cores with evidence of ongoing or recent star formation. In contrast, only about 5 per cent of such galaxies are found in the Virgo cluster \citep{Lis07}. While the frequency of the blue-cored early-type dwarfs in group environments seems to depend on the stage of dynamical evolution of the group, a substantial fraction (more than 30 per cent) of blue-cored early-type dwarfs is also found in nearby groups (e.g., NGC 5846 and NGC 5353/4 groups; \citealt{Mah05}; \citealt{Tul08}). Especially in the NGC 5353/4 group, about two thirds of the early-type galaxies are identified as candidates of blue-cored early-type dwarfs (see \citealt{Tul08}). This suggests that low-density environments are very favorable to the formation and/or survival of blue-cored early-type dwarfs. 

Based on the results of blue-cored early-type dwarfs in the Virgo cluster, several possible scenarios for the formation of these galaxies in cluster environments have been suggested (see \citealt{Lis06}; \citealt{Lis09} for details). Galaxy harassment \citep{Moo96} within the cluster appears to be a possible mechanism to transform the infalling late-type disc galaxies to galaxies with early-type morphology. In this process, blue central region with enhanced density could be formed by funnelled gas into the centre \citep{Moo98} and subsequent star formation.

Ram pressure stripping of dwarf irregular galaxies (dIrrs) by the hot intracluster medium could be responsible for the shaping the blue-cored early-type dwarfs, in which interstellar gas in dIrrs are removed except around the central region of the galaxy. On the other hand, it is also plausible that blue-cored early-type dwarfs could be formed via star formation in central regions of dIrrs triggered by tidal interaction with other galaxies. By several bursts of star formation, dIrrs become galaxies classified as blue compact dwarf galaxies \citep{Dav88}. \citet{Mey14} suggested that evolutionary connections between early-type dwarfs and blue compact dwarfs are possible based on the largely overlapping range of their structural parameters (such as effective surface brightness and radius), for which only the mass-dominating old stellar population of the blue compact dwarfs was considered. In the Ursa Major cluster, more than half of Irrs are HSBs (29/54) that could be potential candidates of blue compact dwarf galaxies.

As hypothesized by \citet{Tul08}, the presence of dwarf galaxies exhibiting star formation at their centre may be less related to direct external influence on the galaxy, but instead more governed by how much of the original gas reservoir remains available when a galaxy enters the cluster potential \citep{Sha84}. In this context, it is noteworthy that \citet{DeR13} recently found that the Fornax cluster dwarf elliptical galaxy FCC046, while having a centrally concentrated young stellar population, is surrounded by a significant gas reservoir ($\sim$10$^{7}$ M$_{\sun}$ of HI gas). This galaxy is located in the outskirts of the Fornax cluster, far from the X-ray halo. This discovery strengthens the argument that accreting gas into the galaxy can be a fuel of central star formation.

\subsection{Environmental effects in a lower density cluster}
It was found that the fraction of massive early-type galaxies (E/S0) and dwarf elliptical galaxies is  higher in a dynamically more evolved system with high velocity dispersion (\citealt{Tre02}; \citealt{Tov04}; \citealt{Tov09}). Especially in the case of the NGC 3992 subgroup, which is the most dynamically evolved Ursa Major cluster subgroup, considering its high frequency of early-type galaxies, morphological transformation to early-type galaxies by interactions and merging between galaxies may have occurred more frequently in the early stages of its dynamical evolution than in the other subgroups.

We find that bright early-type dwarf galaxies ($-18$ $<$ M$_r$ $<$ $-17$ in Fig. \ref{F12}) are absent in the Ursa Major cluster at a significance level of 3$\sigma$. This is based on bootstrap resampling \citep{Bar84} using the luminosity function of galaxy types in the Virgo cluster as reference\footnote{
The bootstrap resampling procedure was repeated 5000 times to generate pseudo-data sets randomly and the statistical uncertainties for each magnitude bin are estimated by running such bootstrap simulations. Out of total 603 early-type galaxies of the Virgo cluster, 33 galaxies are selected as a typical pseudo-data set, which is the same number as early-type galaxies of the Ursa Major cluster. In the resampled pseudo-data sets, having not a single early-type galaxy in the magnitude bin of $-18$ $<$ M$_r$ $<$ $-17$ occurs in less than 0.2 per cent of cases, implying that this gap in the Ursa Major early types is significant at the 3$\sigma$ level.}.
The fact that environmental effects are only of weak to moderate strength is likely to be responsible for the absence of bright early-type dwarf galaxies: no late-type galaxies in this luminosity range were transformed into early types. However, from the analysis of colour gradients of the early-type dwarf galaxies in the Ursa Major cluster (Figs. \ref{F14} and \ref{F15}), the colour in their outer part is as red as those in the Virgo cluster. The outer colour of most blue cored early-type dwarf galaxies in the Ursa Major cluster is within the $\pm$2$\sigma$ region around the red sequence of the Virgo cluster early-type dwarfs.

In contrast to the Virgo cluster environment, ram-pressure stripping as a mechanism regulating galaxy evolution is clearly less relevant in the Ursa Major cluster. The spiral galaxies in the Ursa Major cluster do not show any prominent hint of HI deficiency or gas stripping (see Fig. 1 of \citealt{Ver04}), while many examples for ram pressure stripping are found  in the Virgo cluster (e.g., \citealt{Chu09}). This is also supported by no detection of intracluster medium in the Ursa Major cluster \citep{Ver01a}.

Galaxy harassment, which is a mechanism for transforming late-type galaxies into dEs through frequent high-speed encounters with massive galaxies, prevails in cluster environments \citep{Moo96}. However, this process would be insignificant in groups of relatively low density. We adopt the concept of collision rate described by $R=N\sigma S$ from \citet{May01b} to calculate how insignificant harassment is in the Ursa Major cluster as compared to the Virgo cluster. $N$ is the number density, $\sigma$ is the velocity dispersion of cluster, and $S$ is the cross-section of the target galaxy. The number density $N$ is calculated from the number of galaxies of M$_r$ $<$ $-19$ within 1.5 Mpc of each cluster. The result confirms that the galaxies in the Ursa Major cluster are barely affected by multiple tidal influence with low efficiencies $R_{Virgo}/R_{UMa}\sim27$ for harassment.

In group environments with relatively low velocity dispersion, it has been known that interactions between member galaxies are more frequent (\citealt{Too72}; \citealt{Lav88}; \citealt{Byr90}; \citealt{Mam90}). This is supported by HI surveys of galaxies in the Ursa Major cluster in which a large fraction of galaxies show evidence of tidal interactions (see \citealt{Ver01a}; \citealt{Ver01b}; \citealt{Ver04}; \citealt{Wol13}).  From their extensive HI observations of 43 spiral galaxies, \citet{Ver01a} found that 10 galaxies show clear signs of galaxy-galaxy interactions. Additionally, about half of the sample galaxies also exhibit features of warps or asymmetries in their HI surface density profile of discs. Most recently, from a blind HI survey of the Ursa Major cluster, \citet{Wol13} also found a fraction (33 per cent) of candidates showing tidal interactions. Furthermore, note that all three brightest S0 galaxies (NGC 3998, NGC 4026, and NGC 4111) in the Ursa Major cluster show extended HI filaments, indicative of tidal gas stripping from the disc of nearby companions (\citealt{Ver01b}; \citealt{Ver04}; see also fig. \ref{F12} of \citealt{Wol13} for the extended HI envelope around NGC 4026 and NGC 4111). All of these results provide direct evidence of considerable frequency of interactions between galaxies in the Ursa Major cluster and then support the view that tidal interactions may be the dominant mechanism that transforms late-type dwarf galaxies into early-types in group environments.

In this context, tidally induced star formation of dIrrs and their subsequent evolution should be the most plausible scenario related with the formation of the blue-cored early-type dwarfs in the Ursa Major cluster. In order to compare interaction probabilities, we calculate the perturbation parameter $f$ for the galaxies in the Ursa Major cluster and the Virgo cluster. The $f$ value is defined by \citet{Var04} as
\begin{displaymath}  f = log\left( \frac{F_{ext}}{ F_{int} } \right) =  3 log\left( \frac{R}{ D_{p}} \right) + 0.4 (\textit{m}_{G} - \textit{m}_{p}),\end{displaymath}
where R is the size of the galaxy, D$_P$ is the projected distance between the galaxy and the perturber, and m$_G$ and m$_P$ are the apparent magnitudes of the primary and perturber galaxies, respectively. The galaxy with the largest $f$ value is chosen to be the perturber for the primary galaxy. This parameter means the ratio between the internal and tidal forces acting on the galaxy by a possible perturber. 

The median $f$ values for Virgo and Ursa Major early-type galaxies are $-1.10$ and $-3.22$, with standard deviations of 1.21 and 1.47, respectively. Their ratio is large (10$^{(f_{UMa}-f_{Virgo})}$ $\sim$ 0.008). However, if we assume that interactions between the primary and the perturber start in both clusters at the same time, the average interaction time-scale of galaxies in the Ursa Major cluster will be five times longer than Virgo galaxies since velocity dispersion of Virgo cluster (715 km s$^{-1}$) is larger than that of Ursa Major cluster (151 km s$^{-1}$). While this longer interaction time could make both clusters more similar, it is still not explained that the fraction of blue-cored early-type dwarfs in the Ursa Major cluster is larger than that of Virgo cluster. There is one possibility might solve this problem.

We hypothesize that the blue-cored early-type dwarf galaxies could be considered as possible transitional objects which are on the way to red early-type galaxies from the late-type progenitors (\citealt{Lis06, Lis07}; \citealt{Lis09}; \citealt{Kim10}). Although they are formed via tidal interactions in groups or clusters at the same rate, they have a different evolutionary path as a function of their mass or brightness and the environments where they were formed. In a severe environment like Virgo, less massive galaxies with a shallow potential well lose gas in their centres and will passively evolve to be seen as dE/dS0. In dense environments, only massive galaxies with a deep potential well keep their gas in their centres, although losing it in their outer disk. On the other hand, in a less dense environment like Ursa Major, both massive and less massive galaxies keep their gas in their centre. Consequently, Ursa Major possesses a much larger fraction of blue-cored early-type dwarfs than Virgo.

The above hypothesis is supported by Fig. \ref{F18}. Blue-cored early-type dwarfs in the Virgo cluster are located on the brighter magnitude and higher density region than those in the Ursa Major cluster. We conclude that blue-cored early-type dwarfs are found ubiquitously in all environments including the field (e,g., \citealt{Gu06}) while their frequencies vary with environment and may probe the strength of environment. In a forthcoming paper (Kim et al.\ in preparation.), we will investigate blue-cored early-type dwarf candidates in the field environment in comparison with those in cluster (e.g., Virgo) and group (e.g., Ursa Major) environments, which will allow to probe their formation and evolutionary path.

\section*{Acknowledgments}
We are grateful to the anonymous referee for helpful comments and suggestions that improved the clarity and quality of this paper. We thank Marc A.W. Verheijen for useful discussions. This research was supported by Basic Science Research Program through the National Research Foundation of Korea (NRF) funded by the Ministry of Education, Science and Technology (NRF-2012R1A1B4003097). Support for this work was also provided by the NRF of Korea to the Center for Galaxy Evolution Research (no. 2010-0027910). TL\ was supported within the framework of the Excellence Initiative by the German Research Foundation (DFG) through the Heidelberg Graduate School of Fundamental Physics (grant number GSC 129/1). SK acknowledges support from the National Junior Research Fellowship of NRF(no. 2011-0012618).

\label{lastpage}
\end{document}